\documentclass[fleqn,usenatbib]{mnras}

\usepackage[T1]{fontenc}
   \usepackage{amsmath,amssymb,amsbsy,amsfonts}
   \usepackage{hyperref}
\DeclareRobustCommand{\VAN}[3]{#2}
\let\VANthebibliography\thebibliography
\def\thebibliography{\DeclareRobustCommand{\VAN}[3]{##3}\VANthebibliography}

\usepackage{graphicx}	% Including figure files
   \usepackage{amsmath,amssymb,amsbsy,amsfonts}
\usepackage{comment}
\usepackage{hyperref,graphicx}    

\newcommand*{\figref}[2][]{%
  \hyperref[{#2}]{%
    Figure~\ref*{#2}%
    \ifx\\#1\\%
    \else
      \,#1%
    \fi
  }%
}
\usepackage{tikz}
\usetikzlibrary{arrows,matrix}

\title[Exploring blazars through sonification]{Exploring blazars through
sonification.\\
Visual and auditory insights into multifrequency variability}

\author[G. Magallanes-Guij\'on \& S. Mendoza]{
Gustavo Magallanes-Guij\'on\thanks{E-mail:
gustavo.magallanes.guijon@ciencias.unam.mx (GM)}
and Sergio Mendoza\thanks{
E-mail: sergio@astro.unam.mx (SM)}
\\
Instituto de Astronom\'{\i}a, Universidad Nacional
                 Aut\'onoma de M\'exico, AP 70-264, Ciudad de M\'exico 04510,
                 M\'exico}

\date{Accepted XXX. Received YYY; in original form ZZZ}

\pubyear{2025}

\begin{document}
\label{firstpage}
\pagerange{\pageref{firstpage}--\pageref{lastpage}}
\maketitle

\begin{abstract}
\noindent
Using open astronomical multifrequency databases, we constructed light
curves and developed a comprehensive visualisation and sonification
analysis for the blazars Mrk~501, Mrk~1501, Mrk~421, BL~Lacerta,
AO~0235+164, 3C~66A, OJ~049, OJ~287, and PKS~J2134-0153. This study
employed Musical Instrument Digital Interface (MIDI) and Parameter
Mapping Sonification (PMSon) techniques to generate waveforms,
spectrograms, and sonifications. These representations demonstrate that
data visualisation and sonification are powerful tools for analysing
astronomical objects like blazars, providing insights into their
multifrequency variability. This work highlights how sonification and
visualisation can aid in identifying potential patterns, power
variations, regularities, and gaps in the data. This multimodal approach also underscores the importance of inclusivity in scientific communication, offering accessible methods for exploring the complex behaviour of blazars.
\end{abstract}


%%%%%%%%%%%%%%%%%%%%%%%%%%%%%%%%%%%%%%%%%%%%%%%%%%

%%%%%%%%%%%%%%%%% BODY OF PAPER %%%%%%%%%%%%%%%%%%

%%%%%%%%%%%%%%%%%%%%%%%%%%%%%%%%%%%%%%%%%%%%%%%%%%

\section{Introduction}
\label{introduction}
Data visualisation, often referred to as DataViz, serves as a valuable
approach for identifying and discerning patterns or recurring trends within
the behaviours of social or natural phenomena. It leverages its cognitive
and explanatory capabilities through techniques such as time series,
histograms, infographics, and Venn-Euler
diagrams~\citep{friendly2008brief}. Consequently, graphical
representation stands as a fundamental method for conducting research in
the realms of science and study. Moreover, its ubiquity has made it not
only an analytical tool but also the default and almost universal medium
for communicating information across scientific, educational, and even
everyday contexts, where visualisation often constitutes the first and
most natural choice for conveying knowledge \citep{harris1999}.

DataViz stands as an important methodology for both scientific
communication and research. Nonetheless, it is essential to consider a
couple of factors. Firstly, it is important to recognise that visual
communication is not accessible to individuals who are blind or have
visual impairments (BVI)~\citep{perez2019towards}. Secondly, relying
solely on graphical representations might prove insufficient when
dealing with comprehensive data analysis, which involves exploring
large, multidimensional datasets, detecting subtle correlations,
or interpreting complex temporal and spatial patterns. Therefore,
incorporating supplementary review methods—such as statistical
modelling, sonification, and interactive data exploration—becomes
indispensable to ensure a more complete and robust understanding of the
data \citep{Hearst2011}.

Data sonification can serve not only as a means of inclusivity in scientific communication but also as an additional tool for research~\citep{sawe2020using}. By transforming data into sound, it offers an alternative mode of perception that complements visual representations, allowing for a deeper understanding of complex data.

Building on the limitations of purely visual
approaches, data sonification and sound visualization have emerged as
complementary methodologies for exploring and analyzing complex
datasets. Notably, \citet{Kaper2000} have demonstrated how digital sound
synthesis, via tools such as DIASS (Digital Instrument for Additive
Sound Synthesis), combined with immersive sound visualization in
virtual-reality environments like M4CAVE, can facilitate the
identification of patterns, correlations, and anomalies in
multidimensional scientific data. Their work illustrates that auditory
representations not only enhance inclusivity for blind or visually
impaired (BVI) individuals but also provide all researchers with an
additional perceptual channel, allowing subtle temporal and structural
features in the data to be perceived more effectively.

In this context, \cite{kramer1999sonification}~define data sonification as ``the
transformation of data relations into perceived relations in an acoustic
signal for the purposes of facilitating communication or
interpretation.'' This definition positions sonification as a
supplementary tool that can aid in understanding complex datasets.
Research indicates that the human auditory system is particularly
effective at perceiving information distributed over time and detecting
temporal patterns, often outperforming vision in these tasks
\citep{mcadams1993}. Consequently, sonification proves to be an effective
method for enhancing comprehension of temporal and sequential
information, revealing trends, fluctuations, and correlations in
datasets that might be less apparent in visual representations alone
~\citep{guttman2005hearing}.

Another definition of data sonification, provided by
\cite{hermann2008taxonomy}, characterises it as a technique that takes data
as its input and produces sound signals. In some cases, the generation of these sounds may be influenced by additional stimuli or triggers, which are optional and not strictly necessary for the process. This broad definition considers as sonification any process that uses data to create auditory output if the following conditions are satisfied:

\begin{itemize}
    
\item  The sound reflects objective properties or relations of
the input data.

\item  The transformation is systematic. This means that
there is a precise definition provided of how the data
(and optional interactions) cause the sound to change.

\item  The sonification is reproducible: given the same data
and identical interactions (or triggers), the resulting
sound has to be structurally identical.

\item The system is designed to be reusable and
generalizable, meaning it can be applied to different datasets or used
iteratively to verify consistency with the same dataset.

\end{itemize}

Following~\citet{hermann2008taxonomy}
and~\citet{vogt2008sonification}, this paper adopts the definition of
sonification as the data-dependent generation of sound, provided that
the transformation is systematic, objective, and reproducible, so that
it can be used as a scientific method. This definition is chosen because
it emphasizes the methodological rigor required for sonification to be
considered a reliable tool for scientific analysis, distinguishing it
from more artistic or exploratory uses of sound.

Numerous instances demonstrate the effectiveness of data sonification as a
valuable science communication tool. For the purposes of
this paper, science communication refers to the process of conveying
scientific information and insights to diverse audiences—researchers,
students, and the general public—in a manner that is both understandable
and engaging. Notable examples include its application in depicting
salmon migration patterns ~\citep{hegg2018sound}
and visualising fluctuations in brainwave
activity~\citep{parvizi2018detecting}.Another
compelling case is the Geiger counter, a device used to detect and
measure ionizing radiation. In traditional use, the counter produces
audible clicks corresponding to radiation events, effectively sonifying
the underlying radioactive activity. This auditory representation allows
listeners to perceive variations in radiation intensity over time,
providing both an immediate, intuitive understanding of the data and a
practical tool for real-time monitoring.

In the realm of astronomy, early connections between
sound and scientific data can be traced to~\citet{Jansky1933}, who first
detected radio waves emanating from the centre of the Milky Way and
identified them through the auditory noise in telephone communications.
While \citet{penzias65} three decades later detected the cosmic microwave background
radiation using radio instrumentation,
their work exemplifies how radio data—initially experienced as
noise—would later inspire the development of auditory and sonification
approaches in astronomy \citep{kellermann2020open}.

Another example is that of~\citet{morgan} who took the data of the X-ray
emission of the black hole GRS 1915+105 and translated it into audio signals, allowing us to ``hear'' the accretion disk from it~\cite{masetti2013}. 
Also~\cite{chandra2003} has sonified data and released ``Chandra `Hears' of a black 
hole for the first time''.

\cite{abbott2016observation} documented the simultaneous detection of a
brief signal by two gravitational wave detectors, which resulted from
the fusion of two black holes. The researchers converted the data from
the instant of the collision, which they dubbed a ``chirp,''
 a term chosen because the sonified signal exhibits
auditory similarities to the chirping sounds of birds~\citep[][see
\url{https://youtu.be/QyDcTbR-kEA}]{LIGOchirp}, even
though this sound arises solely from the sonification method and not
from any actual audible event. Furthermore, \cite{Berti2016} published
an article entitled ``The First Sounds of Merging Black Holes'', shedding light on the initial auditory cues of black hole mergers. 

In recent times, the National Aeronautics and Space Administration
(NASA) has unveiled sonification of imagery derived from optical data captured by the Hubble Space Telescope \citep{NASA2022} and X-ray data observed by the Chandra X-ray Observatory (CXO) \citep {chandra2020}.

Recent advances in astronomical sonification have demonstrated the
potential of combining auditory and visual data representations. For
instance, \citet{Tucker2022MNRAS} assessed Astronify \citep{Brasseur2023},
a prototype tool integrated within the MAST archive that enables
sonification of evenly sampled light curves. Their study evaluated
the effectiveness of sonification for signal detection at various
signal-to-noise ratios, showing that auditory representations can aid
in identifying high-SNR transits and complement traditional visual
inspection. More recently, \citet{Huppenkothen2024arXiv} presented the
Sonified Hertzsprung–Russell Diagram, which transforms stellar time
series into sound, encoding physically meaningful features that preserve
astrophysical differences between stars through both visual and auditory
media.  Our work extends these approaches by applying sonification
to multiwavelength light curves of blazars, integrating light curves,
waveforms, and spectrograms to explore temporal evolution across distinct
energy regimes (optical, X-ray, and \(\Gamma\)-ray). This multifrequency
perspective highlights variability patterns in active galactic nuclei and
demonstrates how sonification can support both scientific interpretation
and public engagement in high-energy astrophysics.

In the context of communicating science to the BVI
community, especially in the realm of astrophysics and black holes
(BHs) this work sonifies data from nine blazars: Mrk 501, Mrk 1501, Mrk
421, BL~Lacerta, AO 0235+164, 3C 66A (PKS 0219+428), OJ 049 (PKS
0829+046), OJ 287, and PKS J2134-0153. These objects
have been reported in the literature as exhibiting quasi-periodic
variability in different electromagnetic bands. By sonifying their light
curves, our aim is to explore whether auditory perception can aid in
identifying, characterising, and communicating such periodicities and
variability patterns. Further details are given in
Appendices~\ref{apendice} and~\ref{tablas}.

The selection of blazars analyzed in this study was
based on three main criteria: (i) the availability of long-term,
sampled light curves across multiple wavelengths (radio, optical,
X-ray, and gamma-ray), (ii) documented evidence of significant
variability on different timescales, and (iii) their inclusion in public
monitoring programs that ensure data accessibility and reproducibility.
The chosen sources—such as Mrk~421, Mrk~501, and 3C~273—are among the
most extensively observed and best-characterized blazars in the
literature, known for their pronounced flux variations and strong
multi-band correlations \citep{Abdo2011ApJb}.
These properties make them ideal test cases for exploring how
sonification can enhance the perception and interpretation of temporal
patterns in multifrequency datasets. 

The ensemble of data employed in this study was selected to encompass a
broad range of energy bands (radio, optical, X-ray, and \(\gamma\)-ray)
in order to capture the multifrequency variability characteristic of
blazars. While the temporal coverage is not fully simultaneous across
all bands, the datasets from the \textit{Swift} and \textit{Fermi}
observatories are largely contemporaneous, as both missions were
launched in 2008, providing overlapping X-ray and gamma-ray observations
over extended periods. This temporal alignment allows for meaningful
cross-band comparisons in several sources. For other frequency ranges,
such as optical and radio, the data were chosen to complement these
high-energy observations, offering insights into long-term variability
trends. From a sonification perspective, these differences in time
sampling and frequency response enrich the auditory experience, allowing
users to perceive asynchronous variability patterns, phase shifts,
and multiwavelength correlations through sound.

This sonification serves a dual goal: it enables BVI audiences to directly engage with astrophysical data, perceiving temporal patterns and variability that are typically conveyed visually, while also providing astronomers with an analytical tool to detect trends and anomalies in light curves through auditory perception, thereby complementing conventional visual analysis methods.

Although this study does not aim to definitively demonstrate periodicities,
the auditory representation provides a means to detect potential patterns,
correlate events across different time scales, and explore the complex
dynamics of these astrophysical objects. By demonstrating how sonification
can complement traditional visualizations, this work establishes a
methodological foundation for future research, including systematic
searches for periodicities, transient events, and multiwavelength
correlations in astronomical data. This multimodal approach enables a
more inclusive and perceptually rich exploration of the variability of
blazars for both sighted and visually impaired audiences.

Blazars are extragalactic sources of intense
luminosity, with energy outputs ranging from \(10^{41}\) to \(10^{47}\)
ergs s\(^{-1}\) \citep{Blandford2019}, whose relativistic jets are
oriented close to the observer’s line of sight \citep{Ulrich1997}. This
geometric alignment results in strong Doppler boosting, which amplifies
the observed emission and produces rapid variability across the
electromagnetic (EM) spectrum. Their radiation is
predominantly non-thermal and arises from two main mechanisms:
synchrotron emission, which dominates the radio-to-optical range, and
inverse Compton scattering, responsible for X-ray and gamma-ray photons
\citep{Padovani2012}.Their pronounced variability over a wide range of timescales makes them particularly suitable for this study, where they serve as test cases to assess the potential of sonification both as an analytical tool and as a medium for science communication.

The variability observed at different wavelengths
reflects physical changes in distinct regions of the jet — for instance,
fluctuations in magnetic field strength, particle acceleration, or shock
propagation — leading to complex temporal structures that can range from
minutes to years. Because these variations encode valuable information
about the underlying astrophysical processes, blazars represent ideal
candidates for data sonification. Transforming their multi-band light
curves into sound provides an alternative perceptual channel for
detecting periodicities, correlations, or transient events that might be
less evident through visual inspection alone.

With this background, in
Section~\ref{datagetting} we describe how we obtained the multifrequency data
(radio, optical,
X-ray, and \(\gamma\)-ray). Section~\ref{sonification} shows the
sonification procedure: pre-production, production, and post-production. The
discussion is shown in Section~\ref{discussion} with which we conclude our
study.

\section{Obtaining the data}
\label{datagetting}
\subsection{Radio}
\label{radio}
The radio datasets for this work were obtained from two data bases: the
University of Michigan Radio Astronomy Observatory
(UMRAO)\footnote{\url{https://dept.astro.lsa.umich.edu/datasets/umrao.php}},
and the Astrogeo Very-long-baseline interferometry (Astrogeo-VLBI)
database.

From UMRAO, we obtained long-term monitoring light
curves of total flux density and linear polarization at 4.8, 8.0, and
14.5 GHz measured with the 26 m radio telescope~\citep{Aller1985ApJS}. For
the sonification, data points from the three frequencies were not
averaged or combined interchangeably; instead, they were arranged in
temporal order while preserving their associated frequency tags. This
approach allowed us to create a continuous auditory sequence reflecting
the multifrequency monitoring cadence, where each tone corresponds to an
observation at a specific band. The selected frequencies are those used
in UMRAO’s long-term variability program, optimized for compact radio
sources. This structure made it possible to represent the variability
across frequencies and identify temporal patterns or irregular sampling
intervals through sound.\footnote{\url{https://lsa.umich.edu/astro/facilities/data-sets.html}} The Astrogeo-VLBI data, corresponding to a frequency of 8.4 GHz \citep{Petrov2025}, provides complementary spatial information through very long baseline interferometry.

The Radio Fundamental Catalog (RFC), of the Astro-geo Center, maintained
by L. Petrov\footnote{ \url{http://astrogeo.org/rfc}} provides precise
positions with milli-arcseconds accuracies, maps, and estimates of
correlated flux densities for thousands of compact radio sources. The
mentioned baselines of 1000–8000 km mentioned in the catalog (RFC) correspond to the distances between the
radio telescopes in the very long baseline interferometry (VLBI)
network. Longer baselines provide higher angular resolution, allowing
for finer spatial details of the sources. These datasets
also include light curves provided in tabular format (flux density vs.
time), which are suitable for variability analysis.

\subsection{Optical}
\label{optical}
For optical, the databases used were: American Association of Variable Star Observers (AAVSO)\footnote{\url{https://www.aavso.org/}} and Zwicky Transient Facility (ZTF)\footnote{\url{https://www.ztf.caltech.edu/}}.

From AAVSO, we obtained publicly available photometric light curves of
blazars in different optical filters, contributed by both amateur and
professional astronomers. The data are distributed as tab-delimited
American Standard Code for Information Interchange (ASCII) tables containing Julian Date, magnitude, filter, and error. The optical AAVSO is a public database founded in
1910 and offers observations of variable stars collected and archived
for world-wide access in collaboration with amateur and professional
astronomers. It is
an international organisation of long-term variable star observers who participate
in scientific discovery through variable star astronomy. 

From ZTF, we used time-series photometry in the g and r
bands obtained from the public ZTF data release~\citep{Masci2019}. The
data products consist of calibrated light curves, delivered in tabular
form (magnitude vs. time with uncertainties), which are extracted from
image differencing pipelines applied to wide-field CCD images. The ZTF measurements were arranged in chronological order while preserving their respective filter labels (g and r), ensuring consistency in multi-band variability analysis. The ZTF is a robotic time-domain survey with a 47 square degree ﬁeld with a 600 megapixel camera to scan the entire northern visible sky at rates of \(\sim \) 3760 square degrees/hour to median depths of \(\textsl{g}\) \(\sim \) 20.8 and \(\textsl{r}\) \( \sim \) 20.6 mag (AB, 5\(\sigma\) in 30 sec). ZTF has been studied in the temporal and dynamic sky as near-Earth asteroids, fast-evolving ﬂux transients, and of Galactic variable sources~\citep[e.g.][]{Graham2019PASP}. 

\subsection{X-rays}
For the X-ray light curves,
 we used data from the
Swift database covering energies in the range 0.3–10 keV. The X-Ray
Telescope (XRT) has two key characteristics that are crucial for this
analysis: a low background and a nearly constant point spread function
(PSF) across the field of view. A constant PSF ensures that the spatial
distribution of X-ray photons from the source is measured uniformly
across the detector, reducing systematic effects when constructing light
curves and enabling more reliable detection of flux
variations~\citep{Moretti2005SPIE}. The
selected blazars are relatively X-ray bright, providing sufficient
photon counts to generate meaningful temporal variations for both
visualization and sonification. The Swift database
provides background-subtracted light curves in Flexible Image
Transport System (FITS)
and ASCII table formats, including time, count rate, and errors. These products were directly used in our analysis with
additional reduction.

\subsection{Gamma-rays}

For the gamma-ray regime, we used data from the Fermi Large Area Telescope
(LAT) in the 100 MeV–300 GeV range. Specifically, we used binned light
curves in FITS files from the public Fermi-LAT Light Curve Repository, which provides
fluxes integrated over pre-defined time bins (typically weekly) in
tabular format (flux vs. time with associated errors). These products
are already processed through the LAT standard analysis pipeline
\texttt{FERMITOOLS}\footnote{\url{https://fermi.gsfc.nasa.gov/ssc/data/analysis/software
/}} and
suitable for variability studies. This instrument observes the sky for \( \sim \) 3 hours and 
covers \( \gtrsim \) 20\% of it for each observation. It has a crystal calorimeter for energy measurement, an anti-coincidence detector to distinguish the background of charged particles and high-resolution converter tracker for direction measurement of the incident gamma-rays~\citep{atwood2009large}.

\section{Sonification}
\label{sonification}
Sonifying the data for each of the blazar frequencies was carried out in three stages: pre-production, production and post-production detailed in the next section.

\subsection{Pre-production}
\label{preproduction}
Sonification pre-production mainly relied on a careful
preprocessing of the data in each band. For the radio and optical light
curves, preprocessing involved identifying and handling anomalies such
as spurious flux values (e.g., negative magnitudes, non-physical
outliers exceeding the 3\( \sigma \) level from the local mean), gaps
produced by instrumental failures, and duplicated or corrupted entries.
These cases were flagged automatically and then treated as follows:

\begin{itemize}
\item \textit{Eliminated} if the value was clearly
spurious (e.g., magnitude far outside the instrument’s dynamic range,
corrupted entries).
\item \textit{Corrected} if a small inconsistency could
be addressed (e.g., date formatting or duplicated entries
consolidated).
\item \textit{Transformed} if a re-scaling or
homogenization was required (e.g., converting magnitudes to flux
densities, or aligning different photometric systems).
\end{itemize}

This preprocessing was performed using the \texttt{AWK}
pattern scanning and processing language, which allows filtering,
restructuring, and formatting the data into consistent light curve
tables.

For the X-ray data, a full reduction was necessary. Raw
event files were processed with the
\texttt{HEASoft}\footnote{\url{https://heasarc.gsfc.nasa.gov/docs/software/lheasoft/}}
package, which applies the standard calibration pipelines (e.g., bias
subtraction, exposure corrections, effective area, and deadtime
corrections). The photon event information, together with spacecraft
attitude (position and orientation), is stored in FITS files, which serve as the standard format for
astronomical data exchange and allow reproducibility of the analysis.

For the \(\gamma\)-ray data, reduction was performed using the
\texttt{Fermitools} software suite. The procedure required specifying the
relevant astrophysical parameters of each blazar—Right Ascension (R.A.),
Declination (Dec.), Region of Interest (ROI), and the start and end
times of the data processing—following the standard pipeline described
by~\citet{nacho} and~\citet{gustavo}.

After these preprocessing steps, all datasets were
homogenized to produce multiwavelength light curves in tabular form,
consisting of discrete time measurements (expressed in Modified Julian
Date, MJD), flux or luminosity, and their associated uncertainties,
typically quoted at the 3\( \sigma \) confidence level.

%%%%%%%%%%%%%%%%%%%%%%%%%%%F I G U R E   W A V E F O R M  MRK 501 %%%%%%%%%%%%%%%%%%%%%%%%%%%%%%
\begin{figure*}
  \begin{center}
  \includegraphics[width=8.0cm]{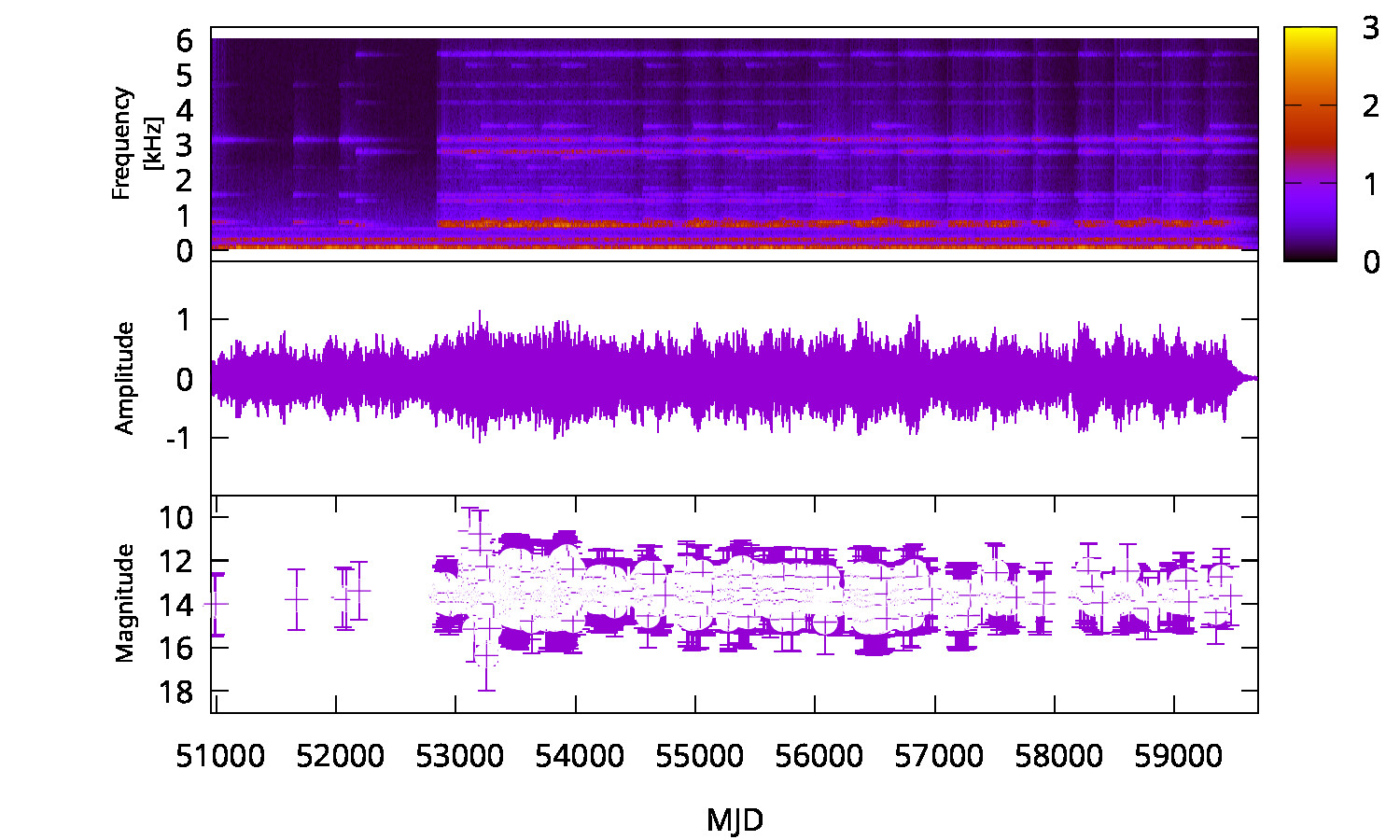} \\
  \includegraphics[width=8.0cm]{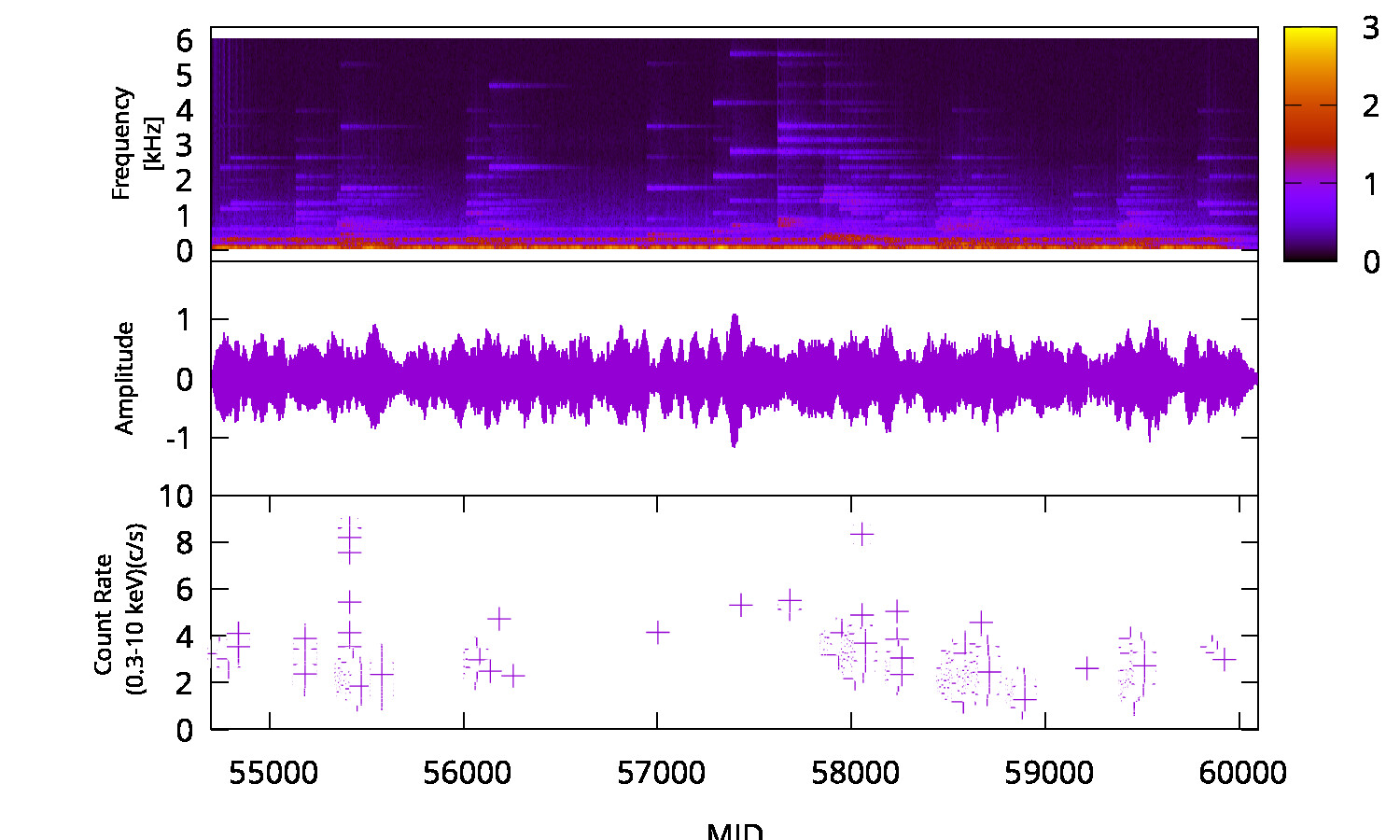}
  \includegraphics[width=8.0cm]{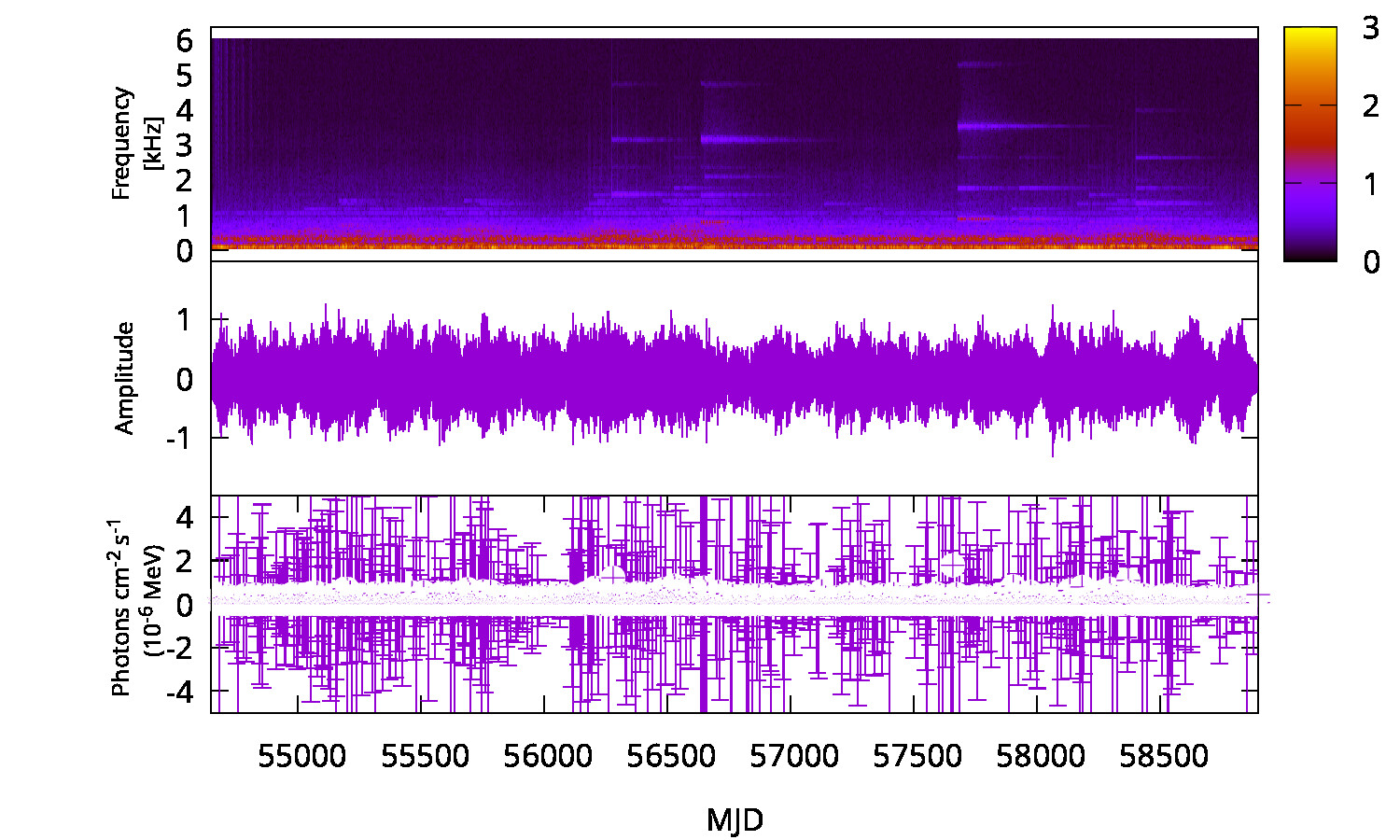} \\
  \end{center}
	\caption{From top to bottom, left to right, the Figure shows panels
	of optical, X-ray and \( \gamma \)-rays  light curves,
	waveforms of the sonification as a function of time, and
	spectrograms of the blazar Mrk~501. The sonification is
	available at 
	\url{https://www.guijongustavo.org/datasonification/mrk501/playlist.html}}.
\label{fig:mrk501}
\end{figure*}
%%%%%%%%%%%%%%%%%%%%%%%%%%%F I G U R E %%%%%%%%%%%%%%%%%%%%%%%%%%%%%%

%%%%%%%%%%%%%%%%%%%%%%%%%%%F I G U R E   W A V E F O R M  MRK 1501 %%%%%%%%%%%%%%%%%%%%%%%%%%%%%%
\begin{figure*}
  \begin{center}
  \includegraphics[width=8.0cm]{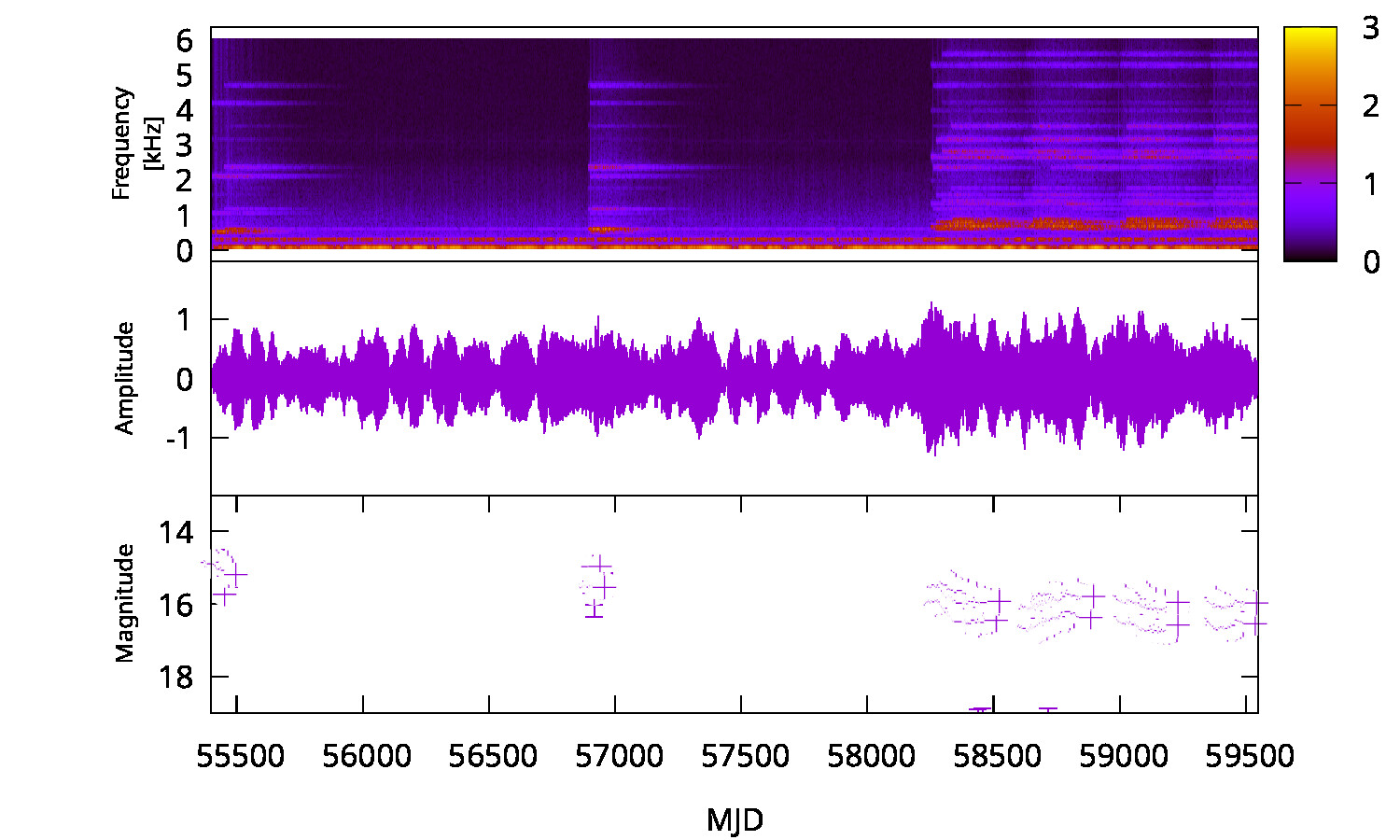} \\
  \includegraphics[width=8.0cm]{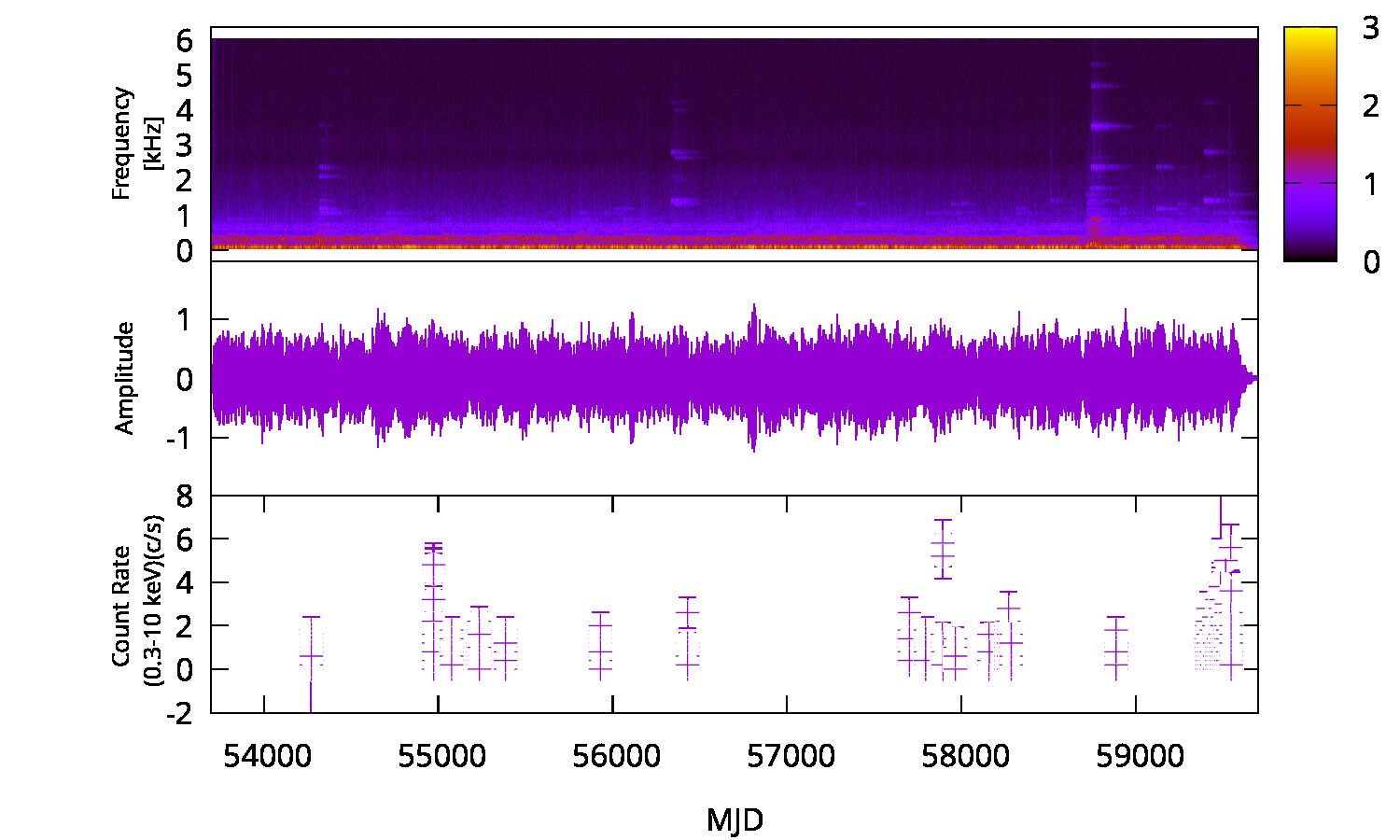}
  \includegraphics[width=8.0cm]{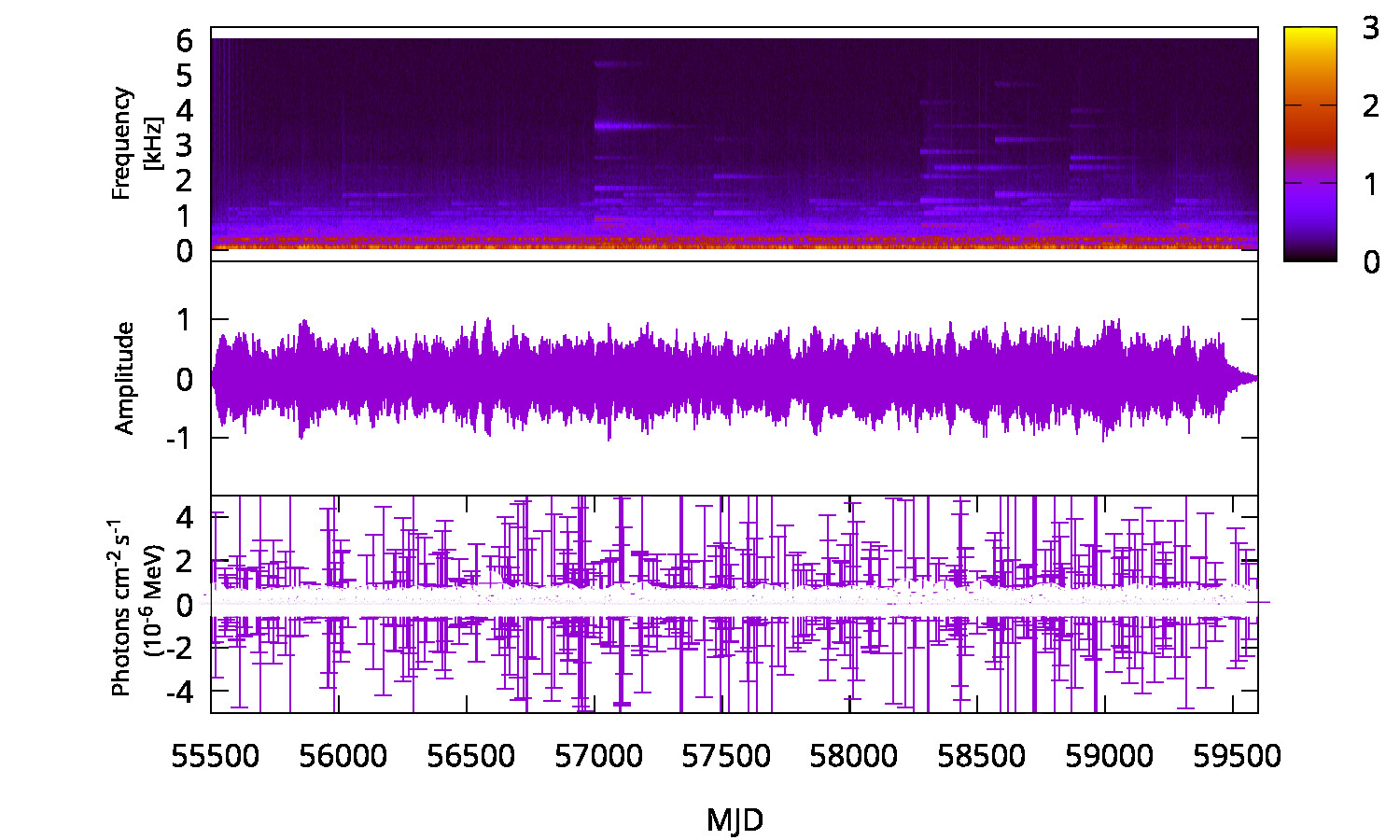} \\
  \end{center}
	\caption{From top to bottom, left to right,the Figure shows panels
	of optical, X-ray and \( \gamma \)-rays  light curves,
	waveforms of the sonification as a function of time, and
	spectrograms of the blazar Mrk~1501. The sonification is
available in \url{https://www.guijongustavo.org/datasonification/mrk1501/playlist.html}}.
\label{fig:wave-mrk1501}
\end{figure*}
%%%%%%%%%%%%%%%%%%%%%%%%%%%F I G U R E %%%%%%%%%%%%%%%%%%%%%%%%%%%%%%

%%%%%%%%%%%%%%%%%%%%%%%%%%%F I G U R E   W A V E F O R M  BL LAC %%%%%%%%%%%%%%%%%%%%%%%%%%%%%%
\begin{figure*}
  \begin{center}
  \includegraphics[width=8.0cm]{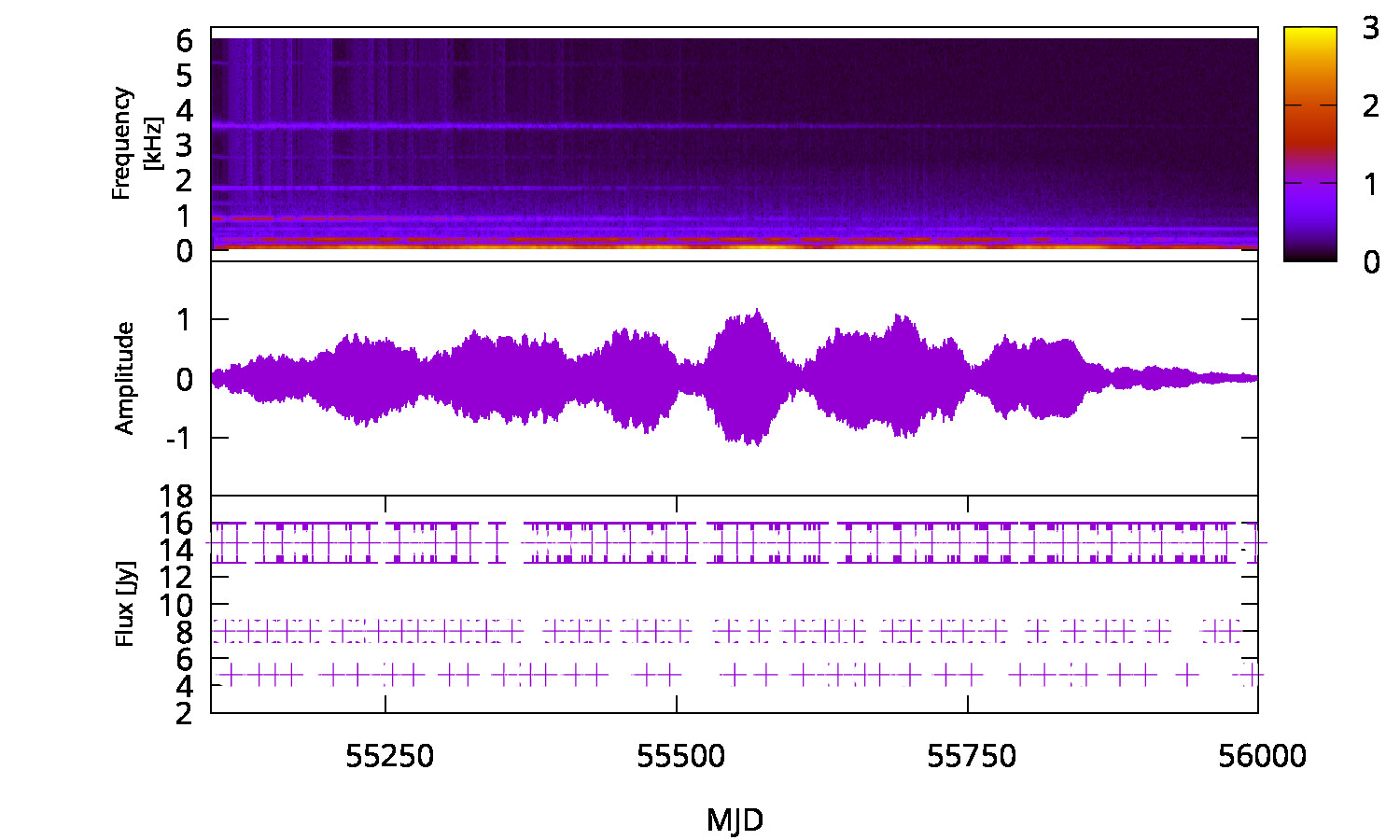}
  \includegraphics[width=8.0cm]{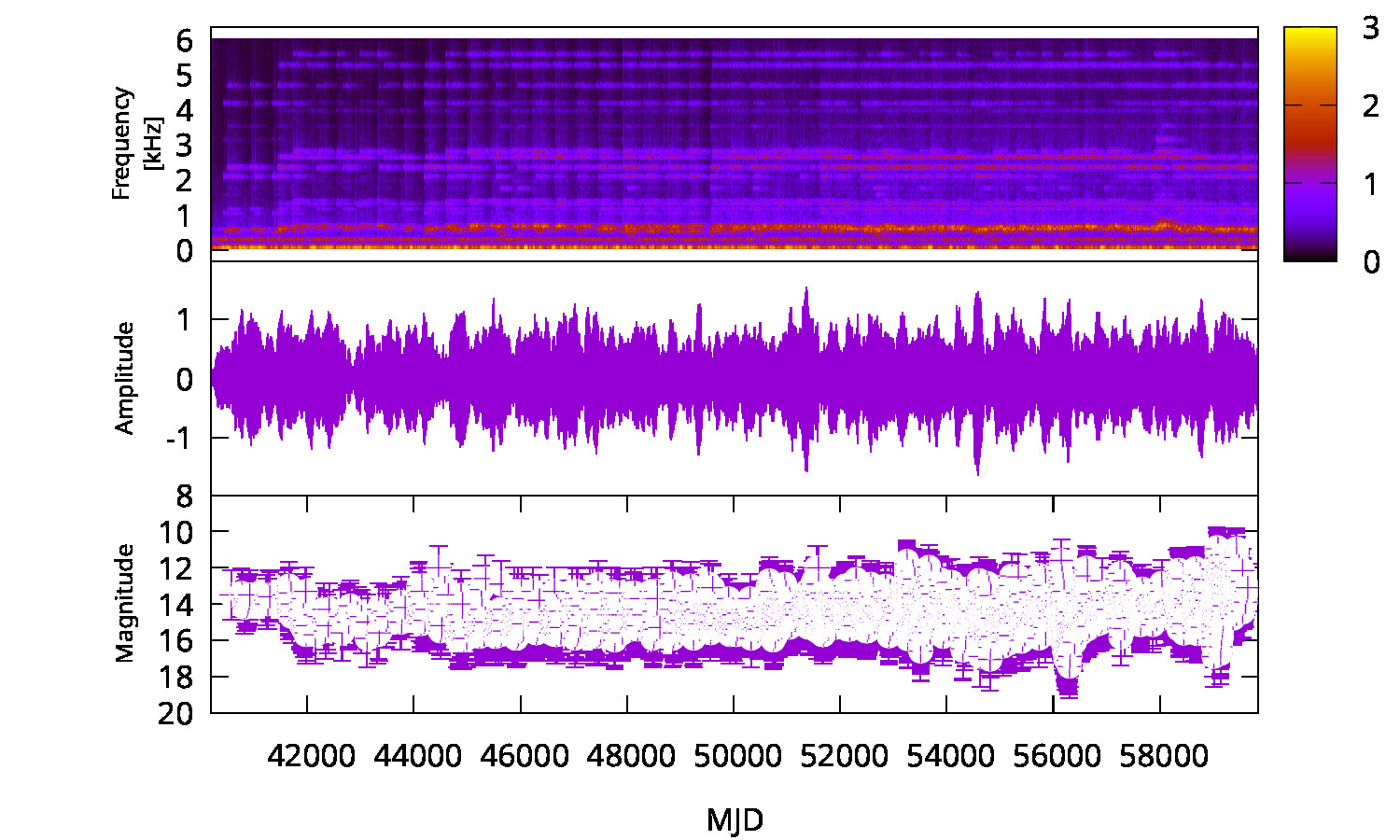} \\
  \includegraphics[width=8.0cm]{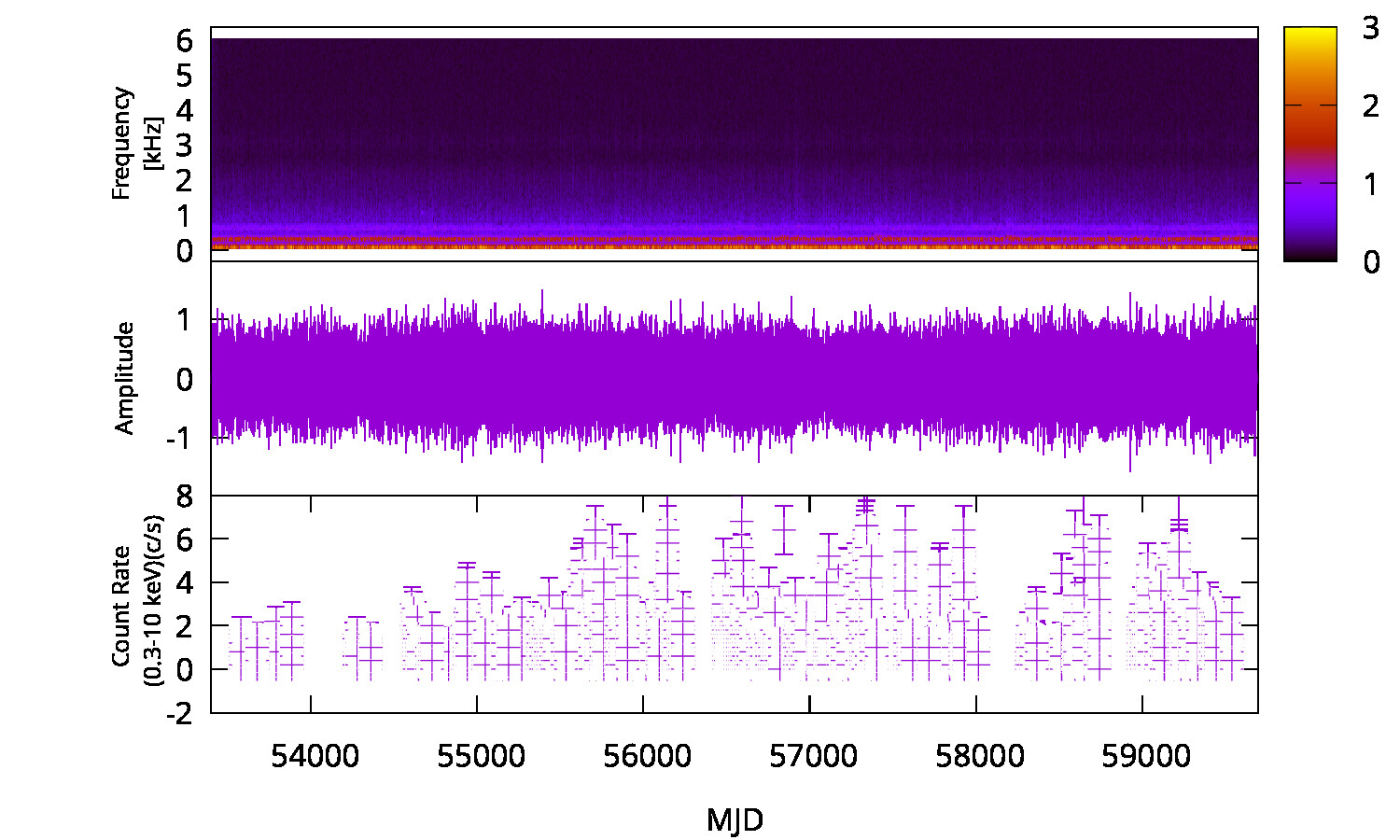}
  \includegraphics[width=8.0cm]{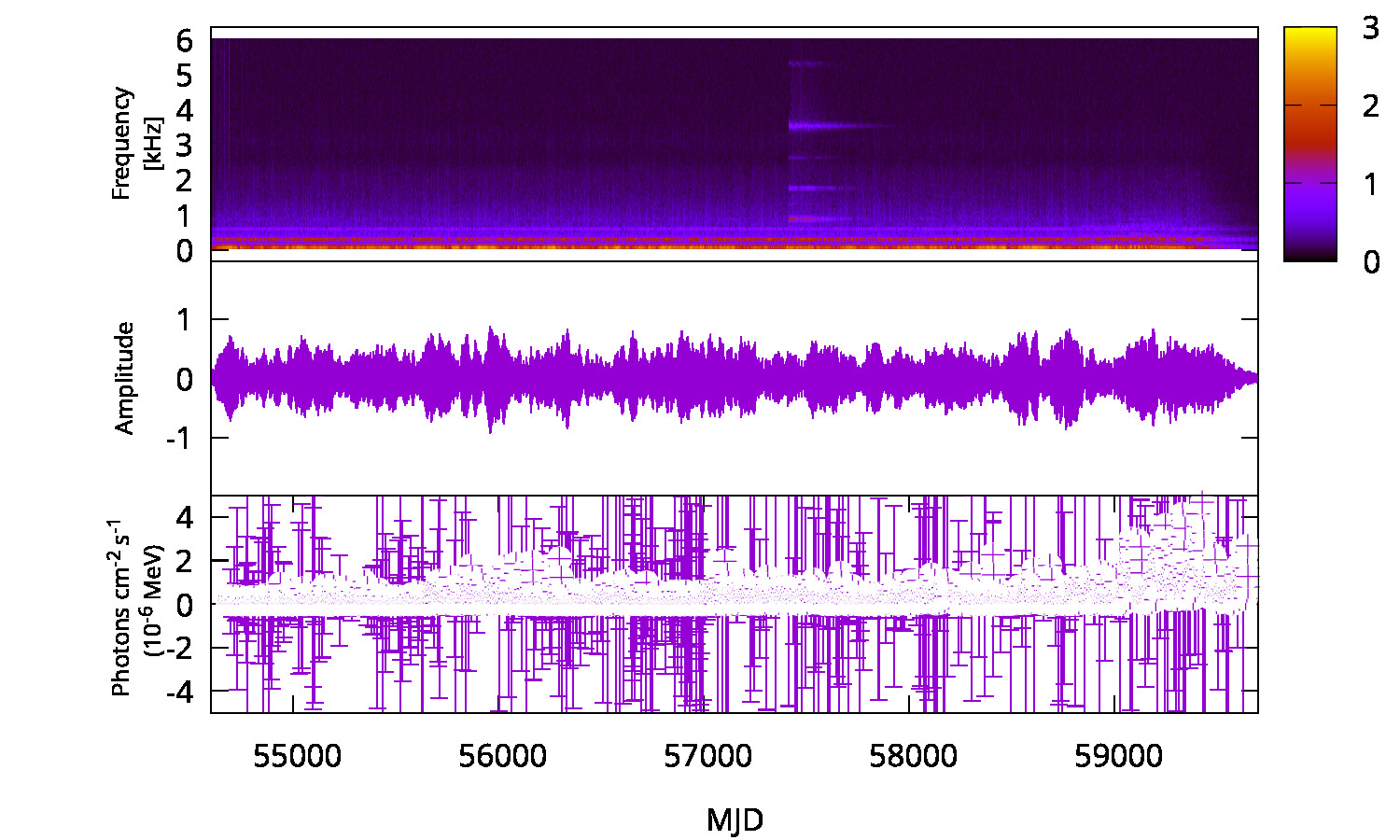} \\
  \end{center}
\caption{From left to right, top to bottom, the Figure shows panels
	of radio, optical, X-ray and \( \gamma \)-rays  light curves,
	waveforms of the sonification as a function of time, and
	spectrograms of the blazar BL Lacerta. The sonification is
available in \url{https://www.guijongustavo.org/datasonification/bllacertae/playlist.html}}.
\label{fig:wave-bllac}
\end{figure*}
%%%%%%%%%%%%%%%%%%%%%%%%%%%F I G U R E %%%%%%%%%%%%%%%%%%%%%%%%%%%%%%
 
%%%%%%%%%%%%%%%%%%%%%%%%%%%F I G U R E   W A V E F O R M  AO 0235+164 %%%%%%%%%%%%%%%%%%%%%%%%%%%%%%
\begin{figure*}
  \begin{center}
  \includegraphics[width=8.0cm]{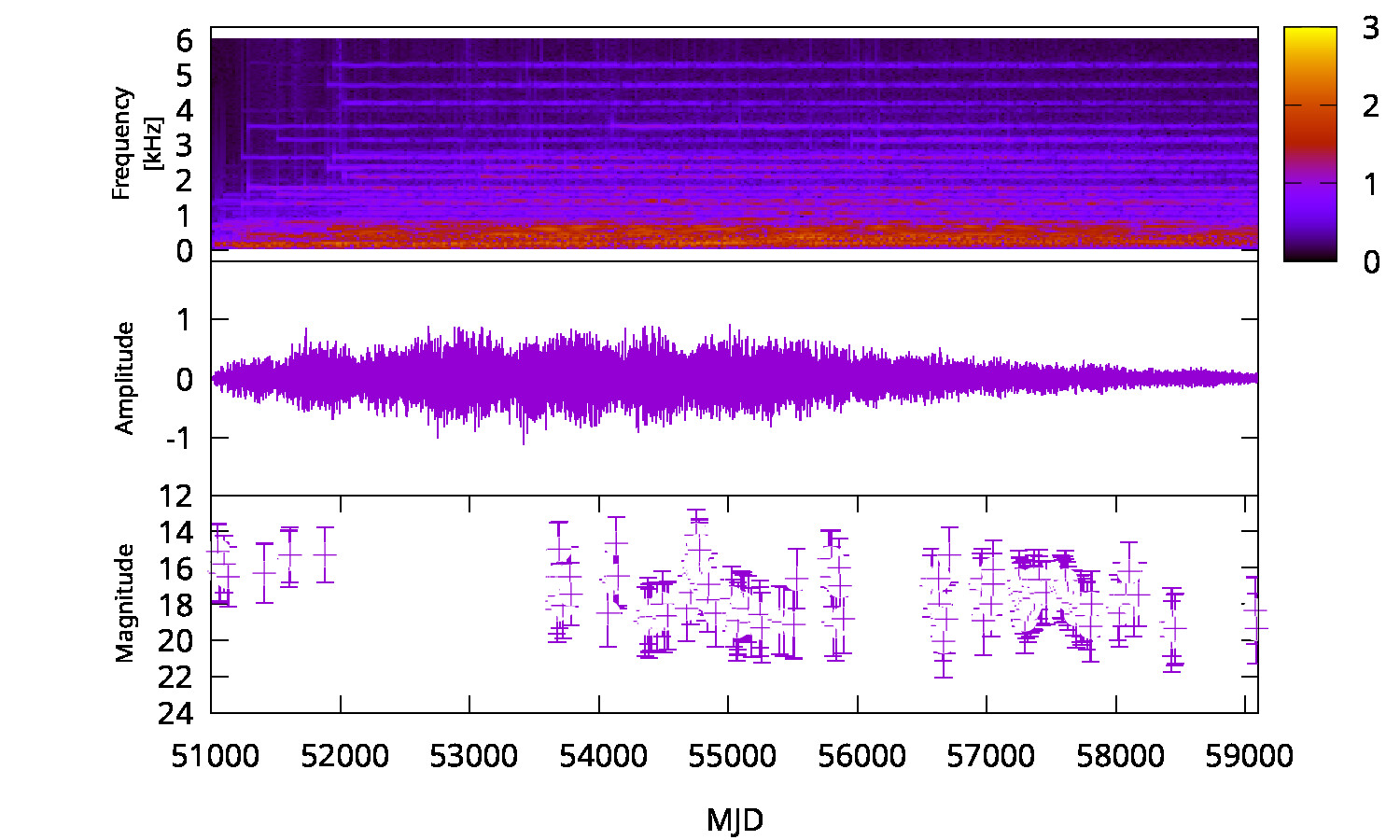} \\
  \includegraphics[width=8.0cm]{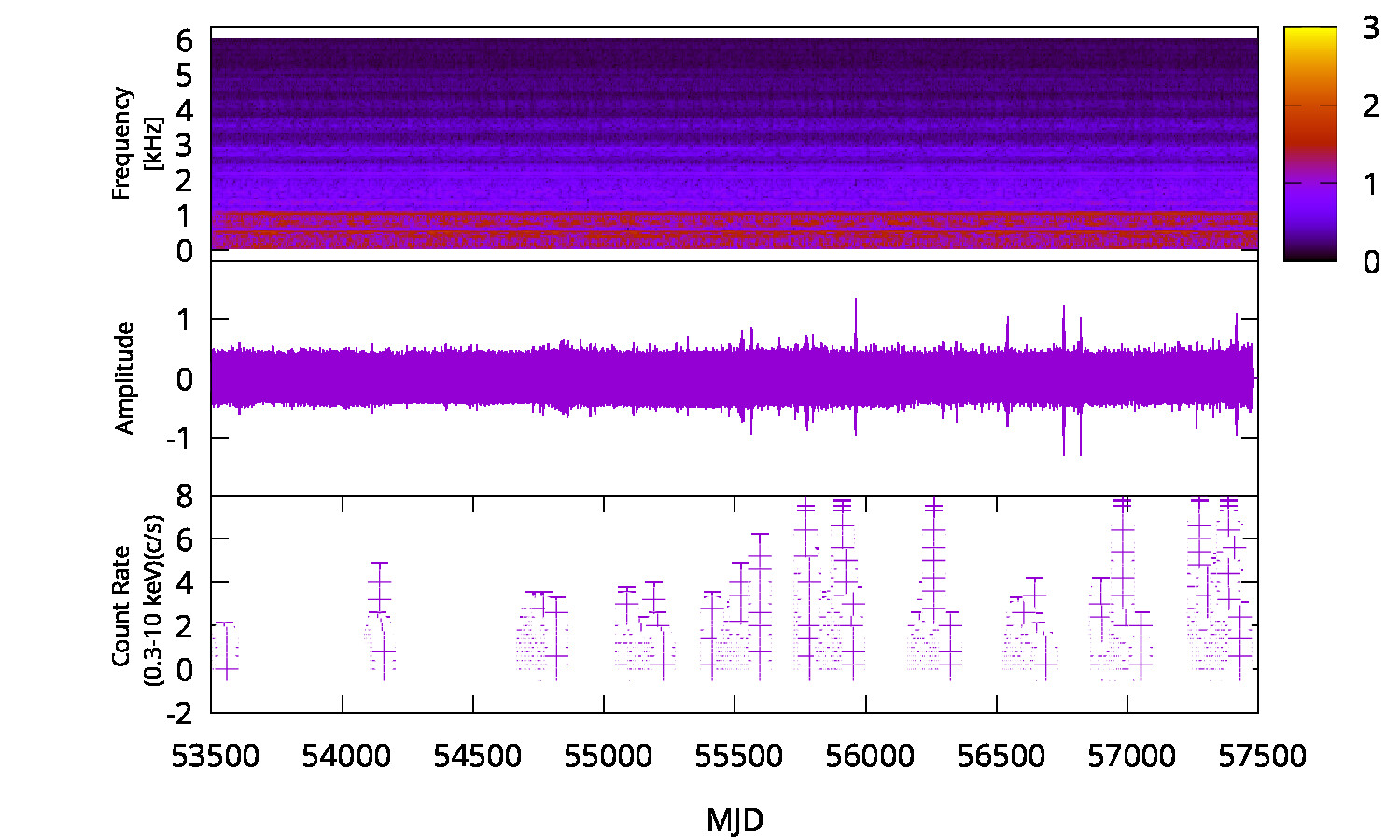}
  \includegraphics[width=8.0cm]{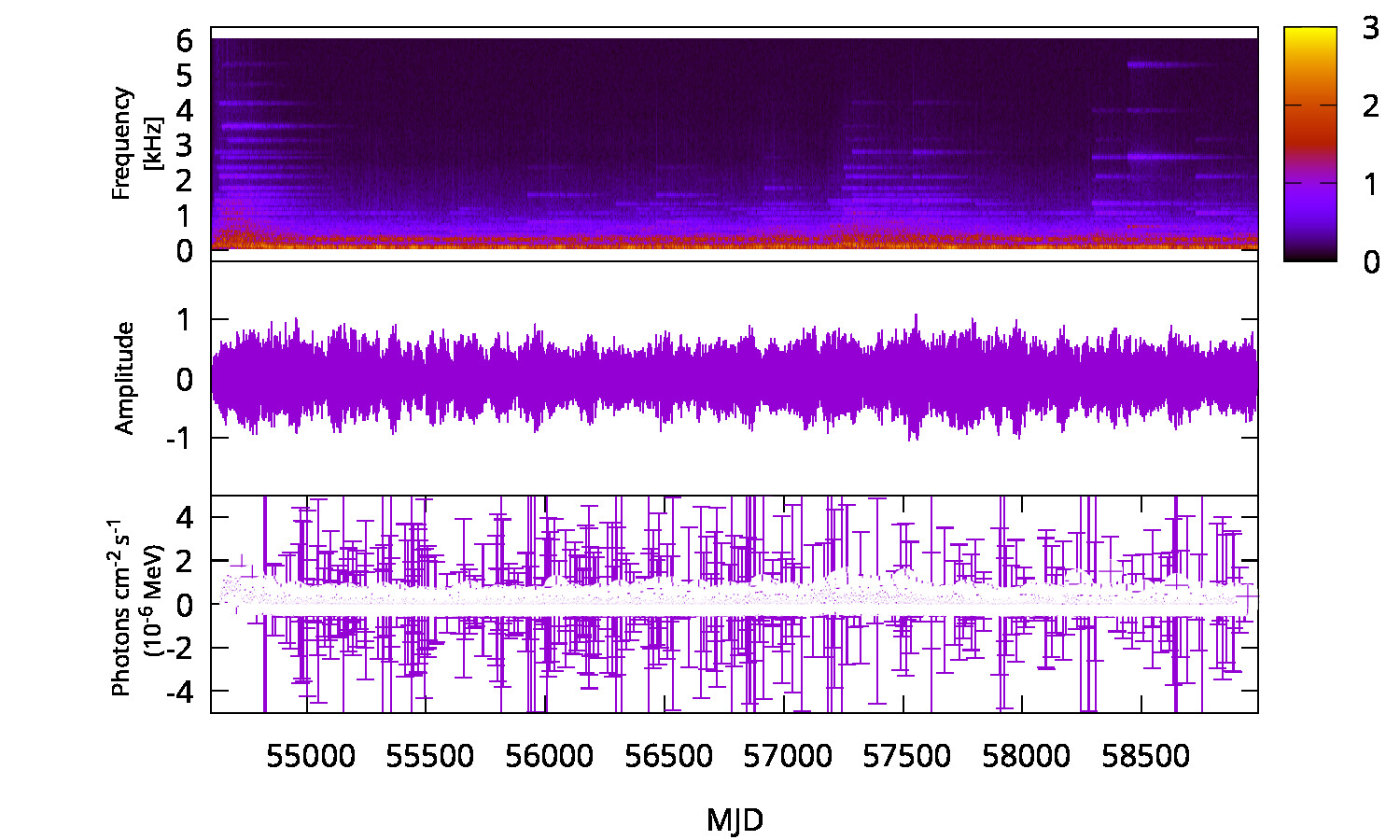} \\
  \end{center}
\caption{From top to bottom, left to right, the Figure shows panels
	of optical, X-ray and \( \gamma \)-rays  light curves,
	waveforms of the sonification as a function of time, and
	spectrograms of the blazar AO 0235+164. The sonification is
available in \url{https://www.guijongustavo.org/datasonification/ao0335/playlist.html}}.
\label{fig:wave-AO0235+164}
\end{figure*}
%%%%%%%%%%%%%%%%%%%%%%%%%%%F I G U R E %%%%%%%%%%%%%%%%%%%%%%%%%%%%%%

%%%%%%%%%%%%%%%%%%%%%%%%%%%F I G U R E   W A V E F O R M  3C66A %%%%%%%%%%%%%%%%%%%%%%%%%%%%%%
\begin{figure*}
  \begin{center}
  \includegraphics[width=8.0cm]{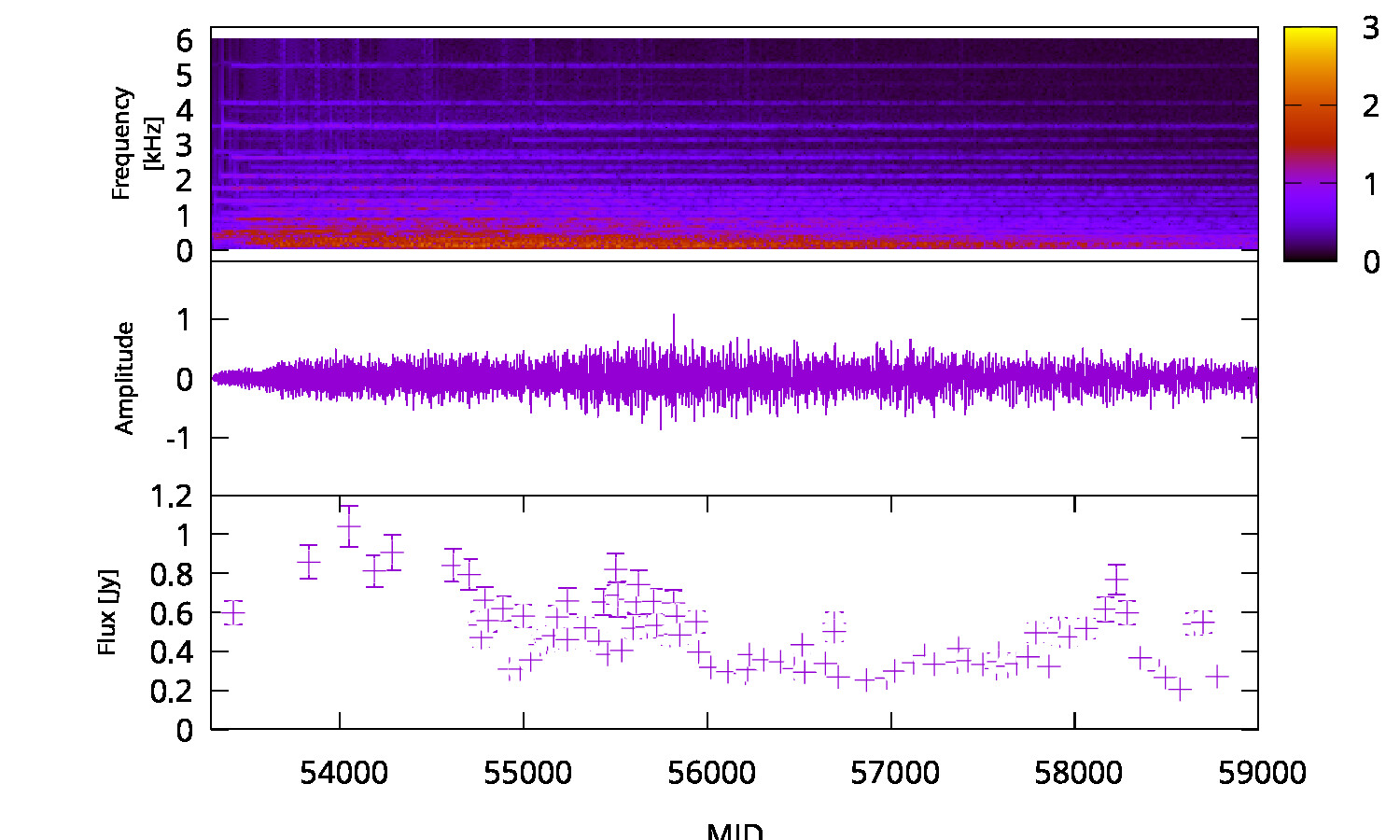}
  \includegraphics[width=8.0cm]{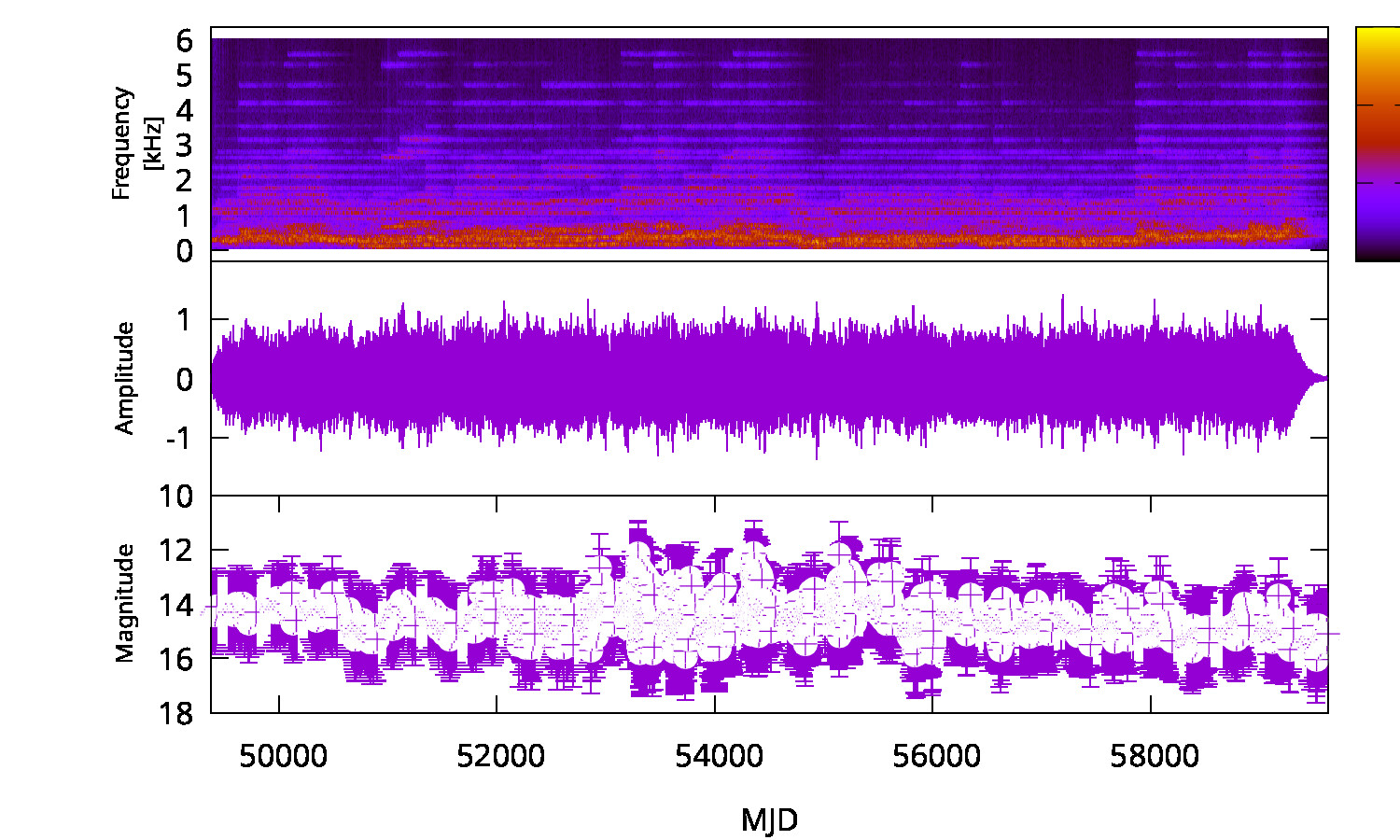} \\
  \includegraphics[width=8.0cm]{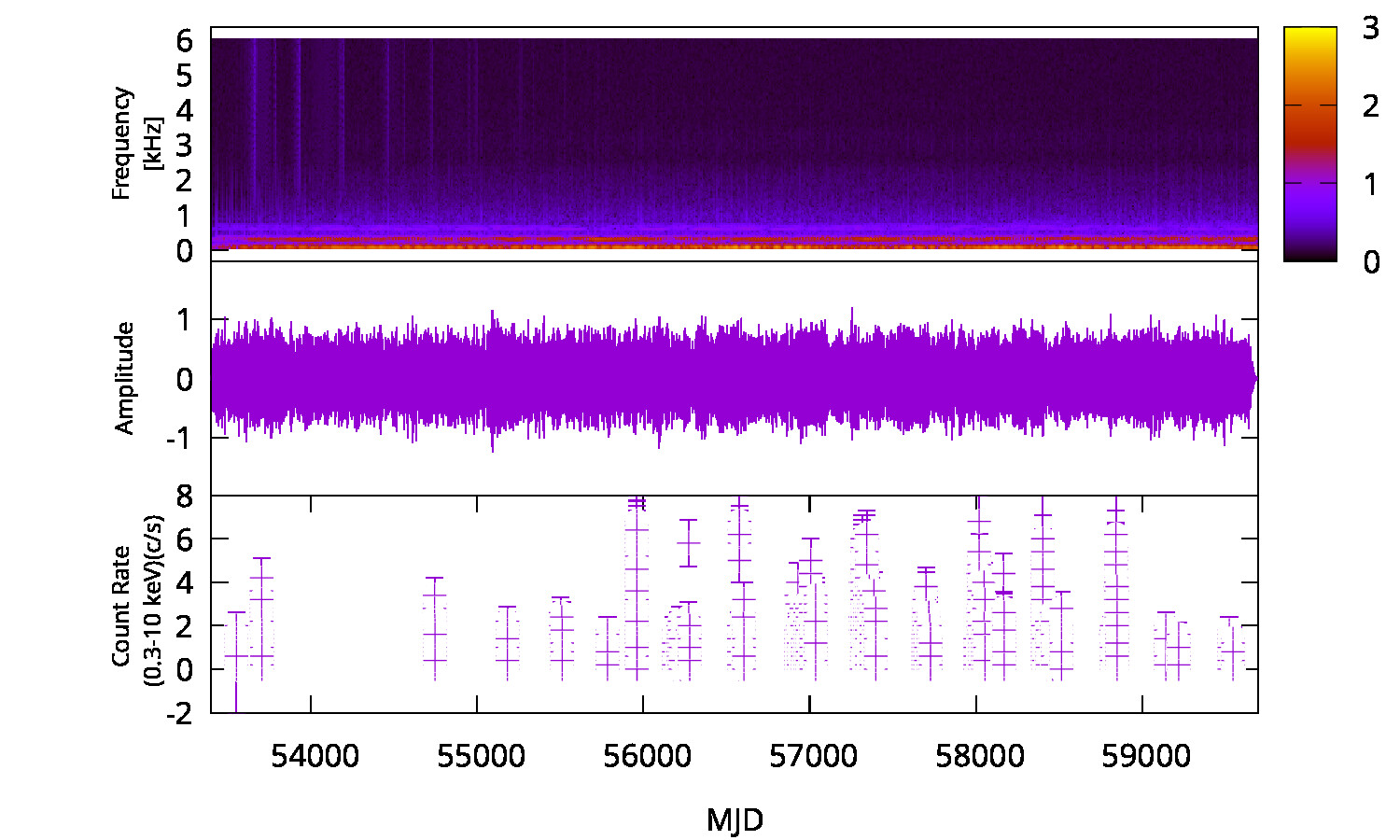}
  \includegraphics[width=8.0cm]{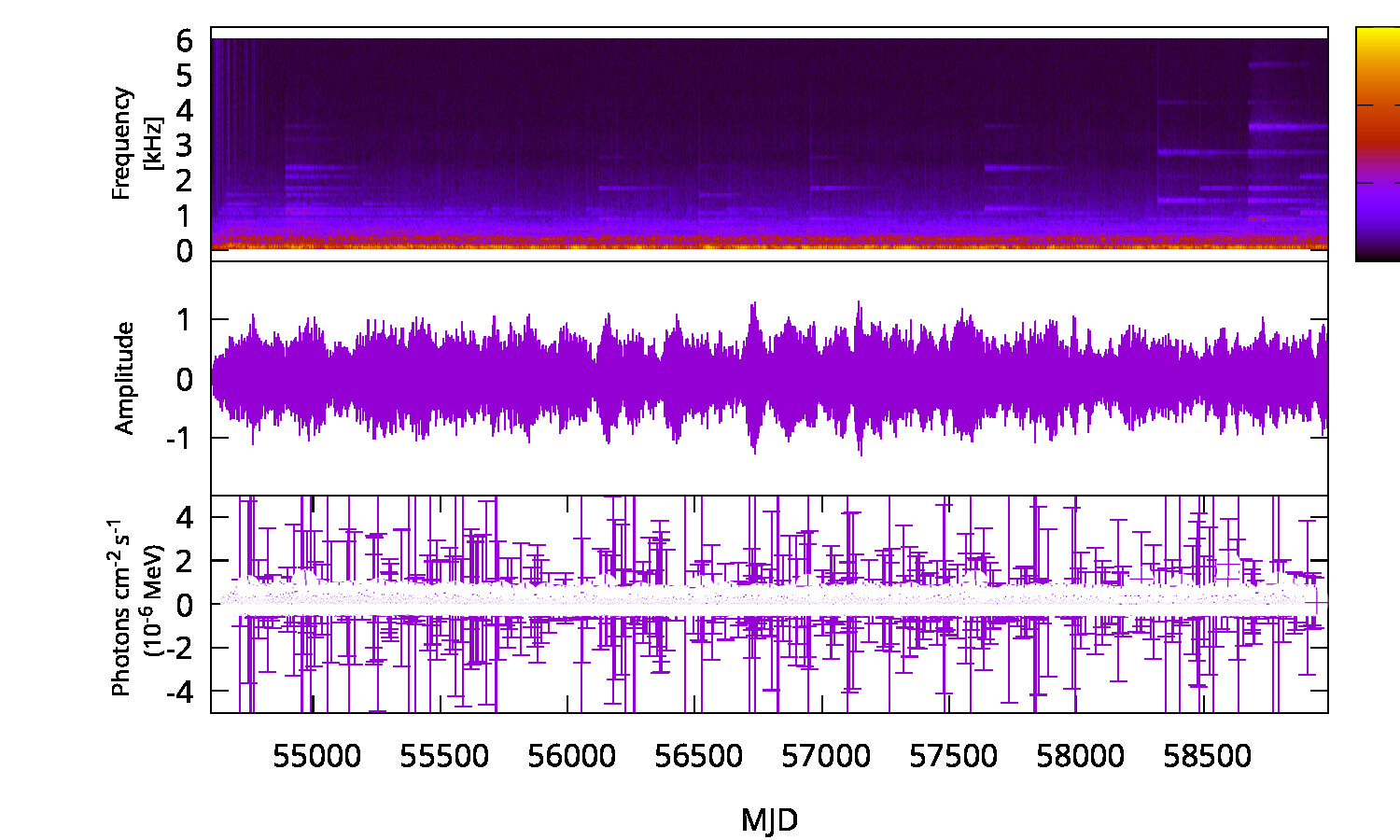} \\
  \end{center}
\caption{From left to right, top to bottom, the Figure shows panels
	of radio, optical, X-ray and \( \gamma \)-rays  light curves,
	waveforms of the sonification as a function of time, and
	spectrograms of the blazar 3C~66A. The sonification is
available in \url{https://www.guijongustavo.org/datasonification/3c66a/playlist.html}}.
\label{fig:wave-3C66A}
\end{figure*}
%%%%%%%%%%%%%%%%%%%%%%%%%%%F I G U R E %%%%%%%%%%%%%%%%%%%%%%%%%%%%%%

%%%%%%%%%%%%%%%%%%%%%%%%%%%F I G U R E   W A V E F O R M  OJ 049 %%%%%%%%%%%%%%%%%%%%%%%%%%%%%%
\begin{figure*}
  \begin{center}
  \includegraphics[width=8.0cm]{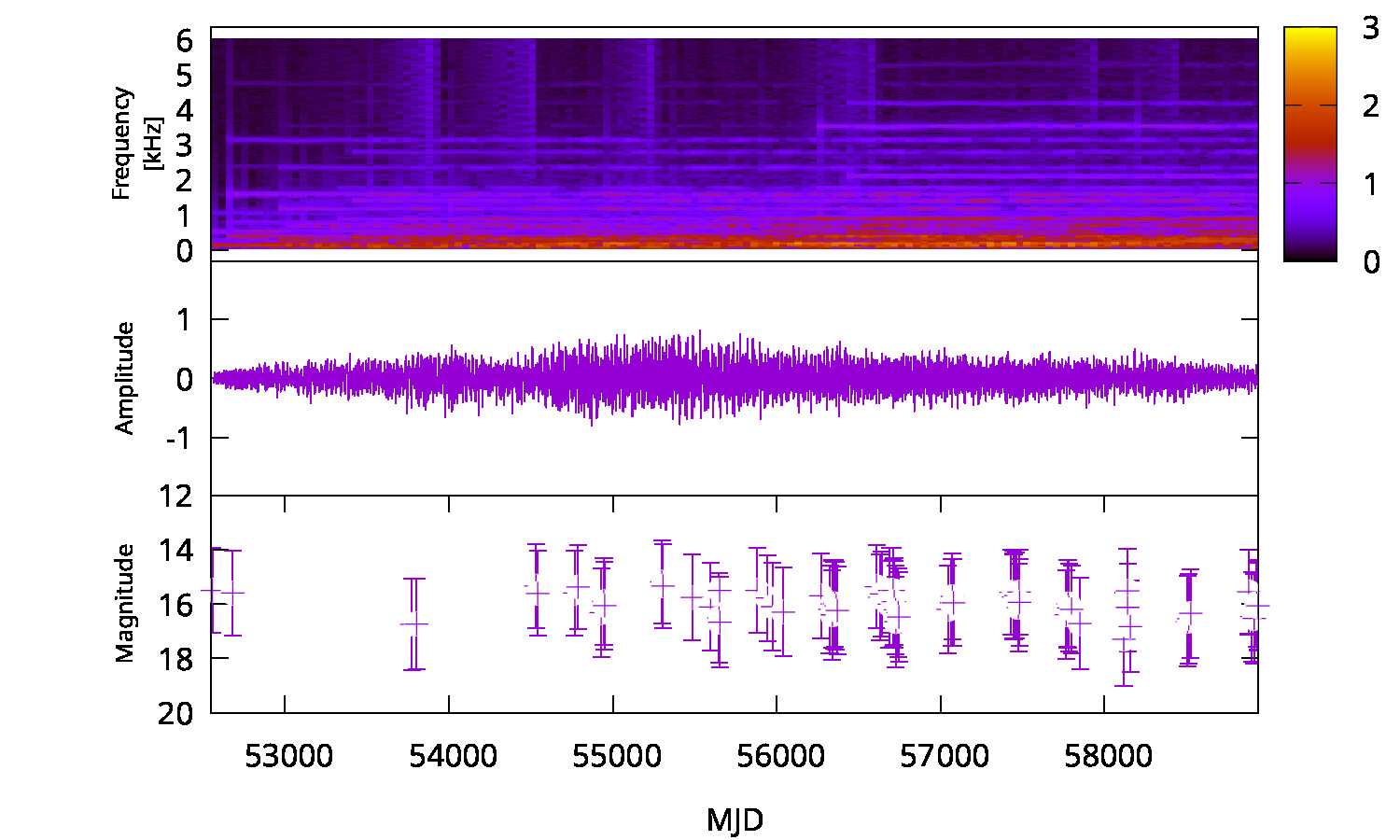} \\
  \includegraphics[width=8.0cm]{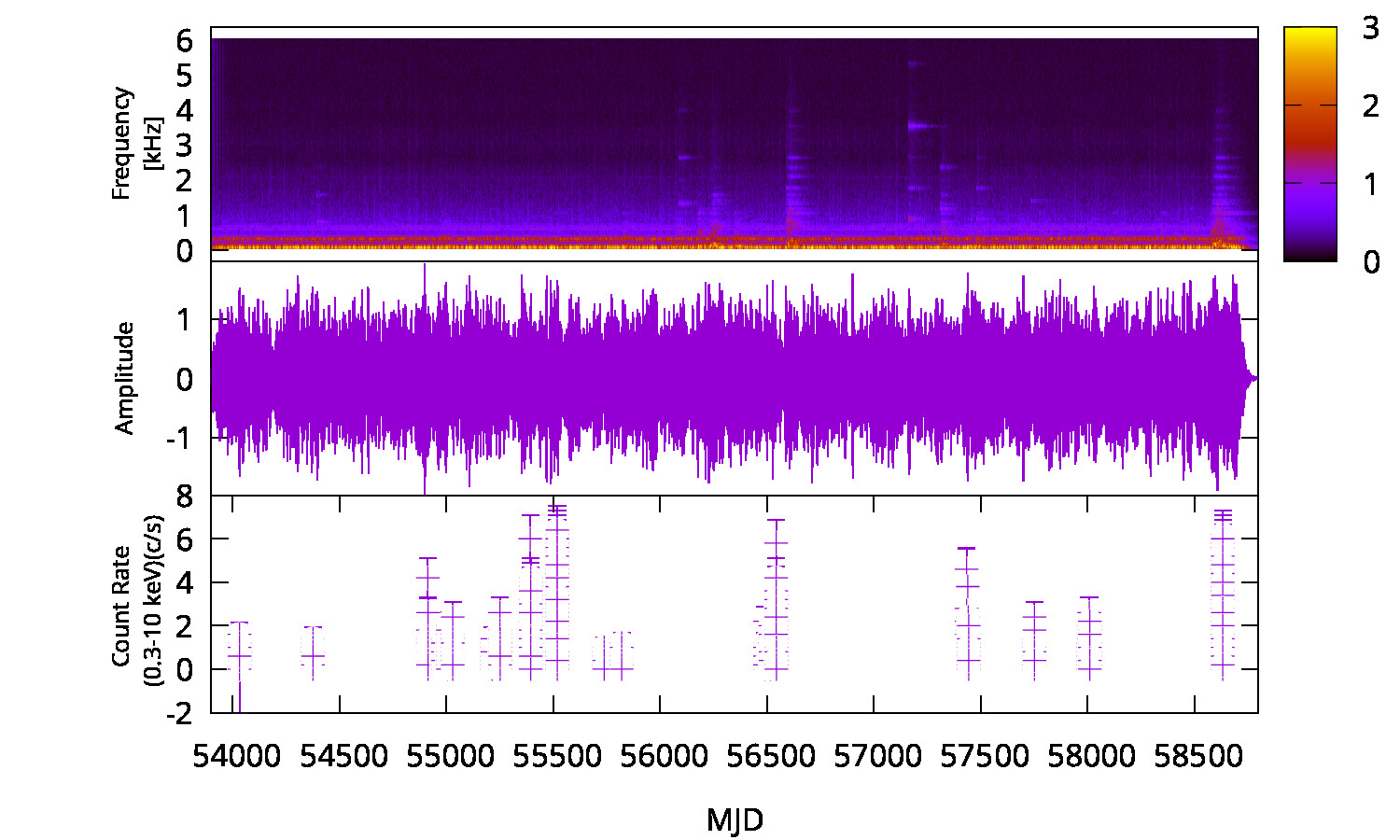}
  \includegraphics[width=8.0cm]{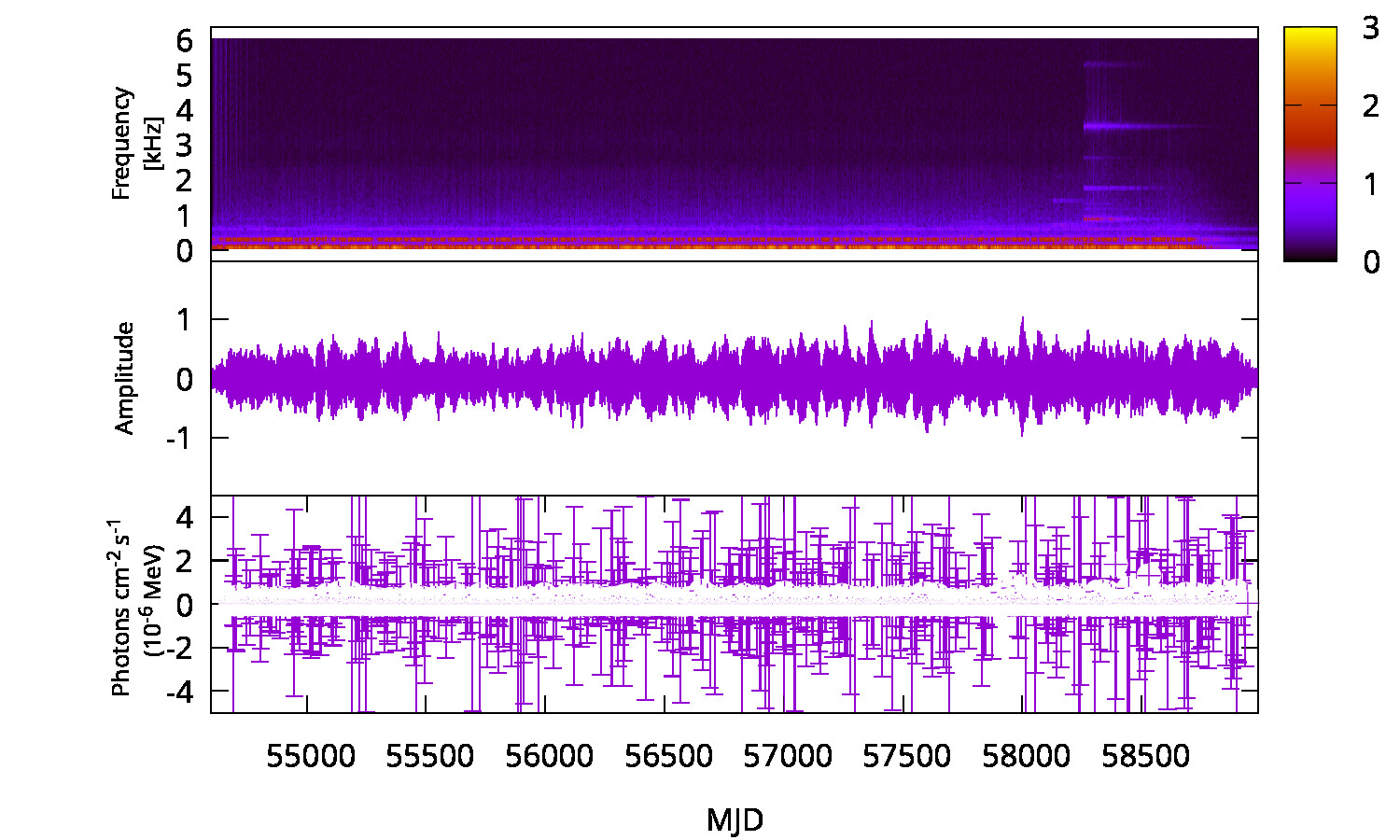} \\
  \end{center}
\caption{From top to bottom, left to right, the Figure shows panels
	of optical, X-ray and \( \gamma \)-rays  light curves,
	waveforms of the sonification as a function of time, and
	spectrograms of the blazar OJ~049. The sonification is
 available in \url{https://www.guijongustavo.org/datasonification/oj049/playlist.html}}.
\label{fig:wave-OJ049}
\end{figure*}
%%%%%%%%%%%%%%%%%%%%%%%%%%%F I G U R E %%%%%%%%%%%%%%%%%%%%%%%%%%%%%%

%%%%%%%%%%%%%%%%%%%%%%%%%%%F I G U R E   W A V E F O R M  J2134-0153 %%%%%%%%%%%%%%%%%%%%%%%%%%%%%%
\begin{figure*}
  \begin{center}
  \includegraphics[width=8.0cm]{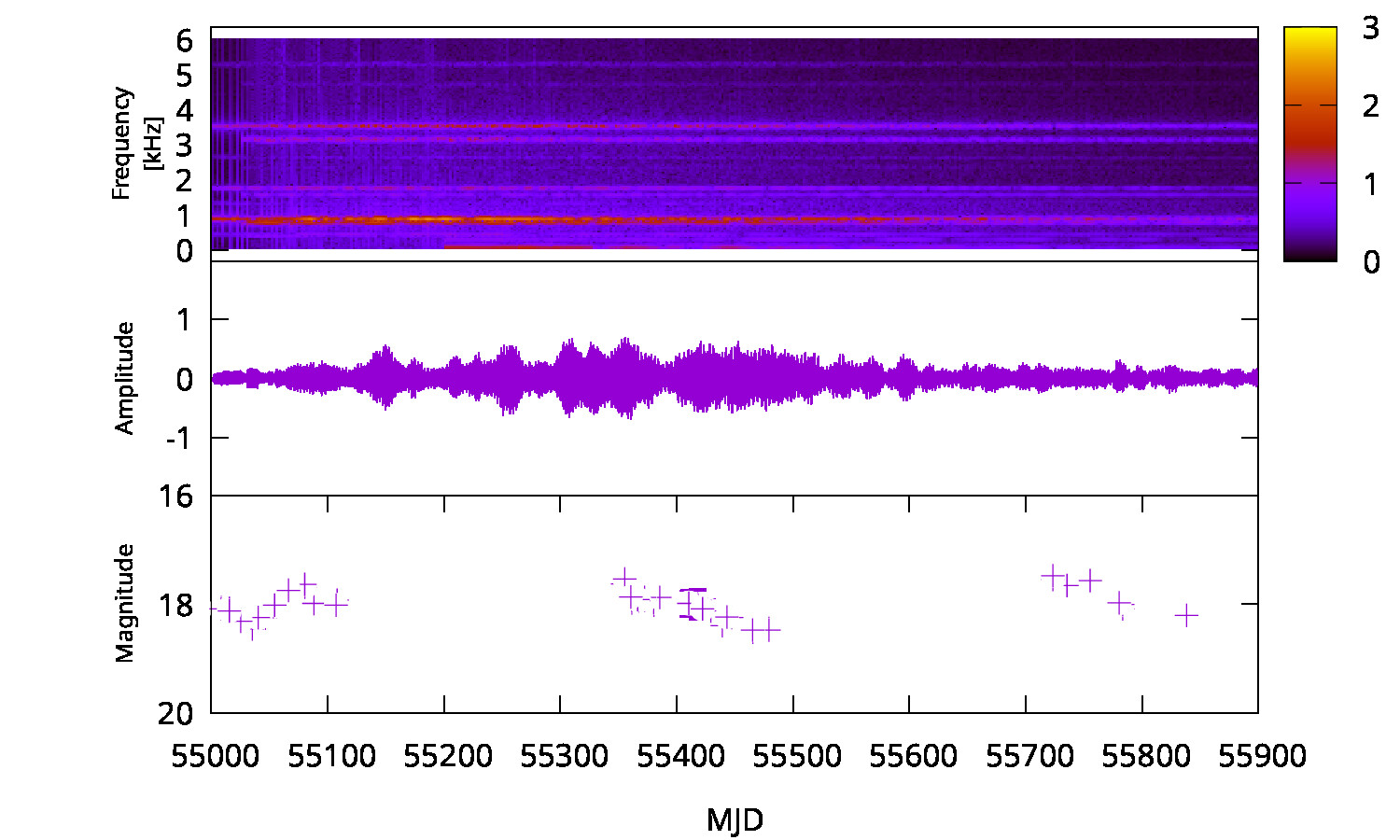} \\
  \includegraphics[width=8.0cm]{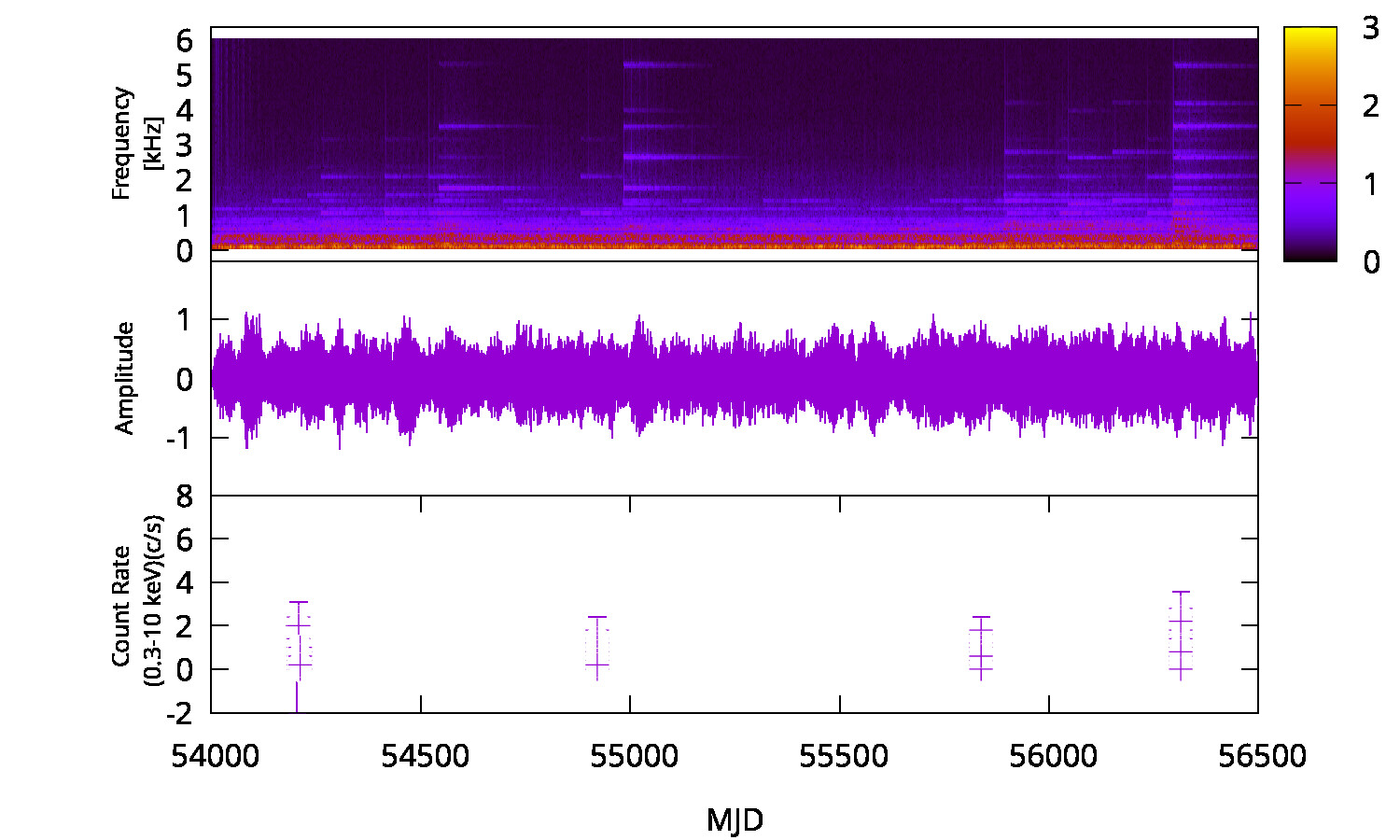}
  \includegraphics[width=8.0cm]{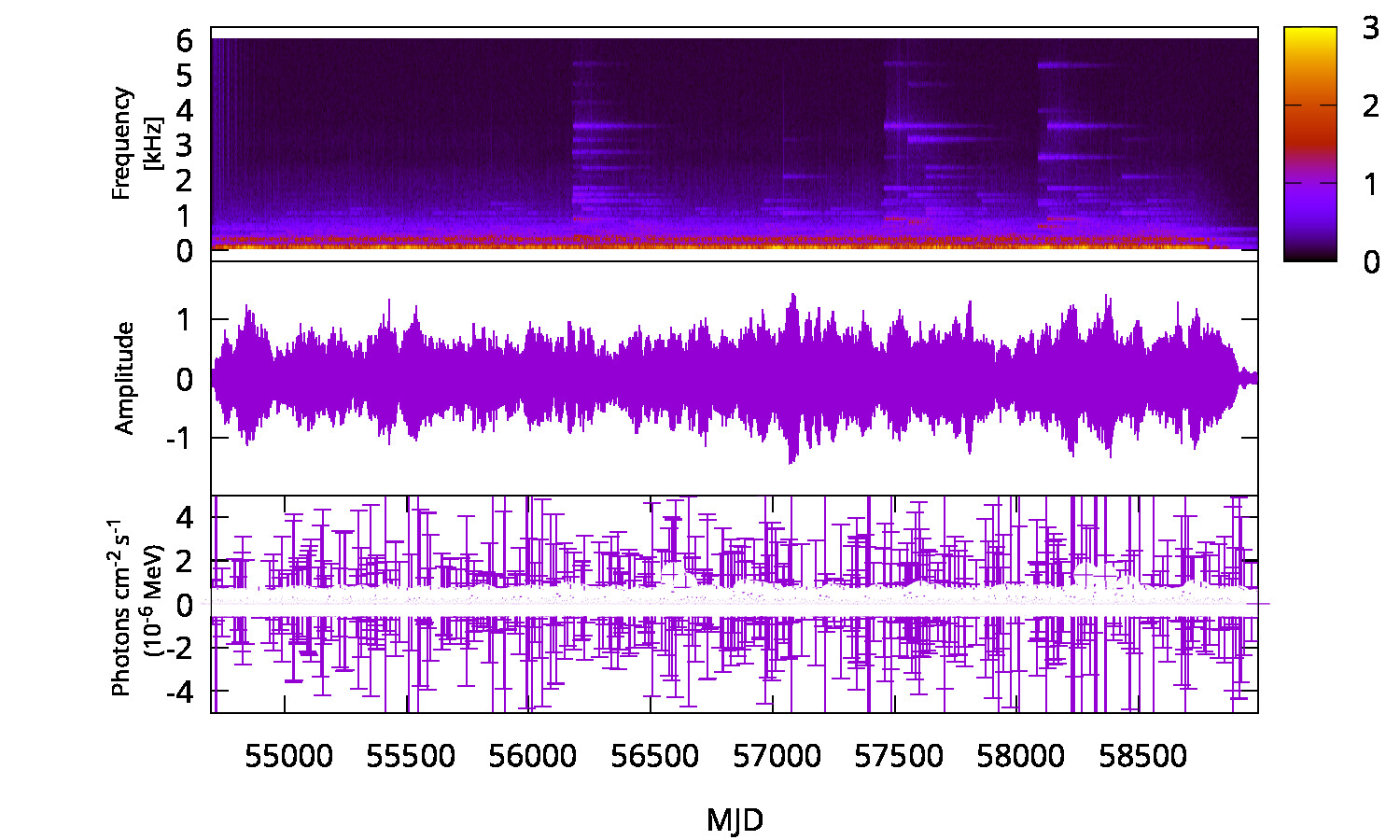} \\
  \end{center}
\caption{From top to bottom, left to right, the Figure shows panels
	of optical, X-ray and \( \gamma \)-rays  light curves,
	waveforms of the sonification as a function of time, and
	spectrograms of the blazar J2134-0153. The sonification is
 available in \url{https://www.guijongustavo.org/datasonification/j2134/playlist.html}}.
\label{fig:wave-J2134-0153}
\end{figure*}
%%%%%%%%%%%%%%%%%%%%%%%%%%%F I G U R E %%%%%%%%%%%%%%%%%%%%%%%%%%%%%%

%%%%%%%%%%%%%%%%%%%%%%%%%%%F I G U R E   W A V E F O R M  MRK421 %%%%%%%%%%%%%%%%%%%%%%%%%%%%%%
\begin{figure*}
  \begin{center}
  \includegraphics[width=8.0cm]{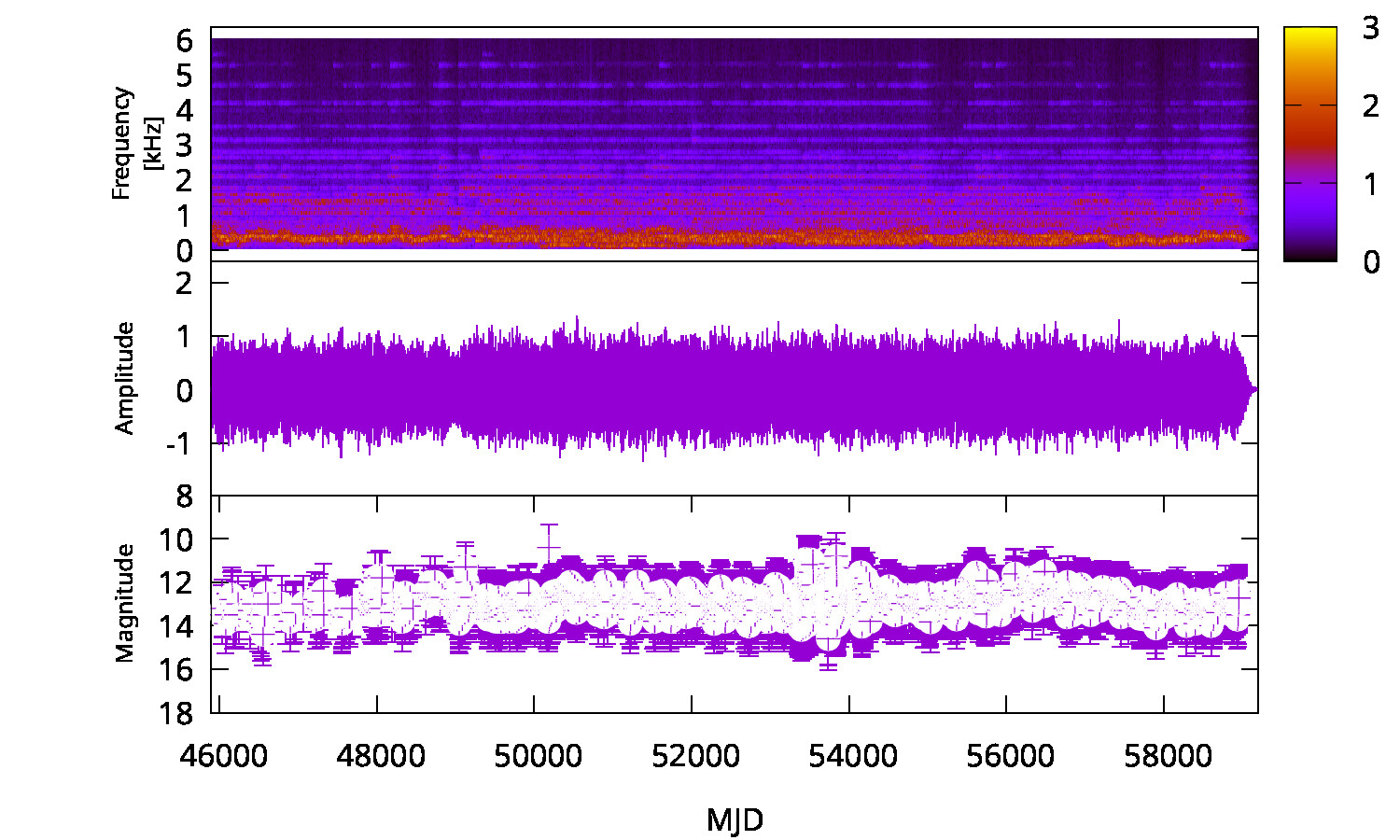} \\
  \includegraphics[width=8.0cm]{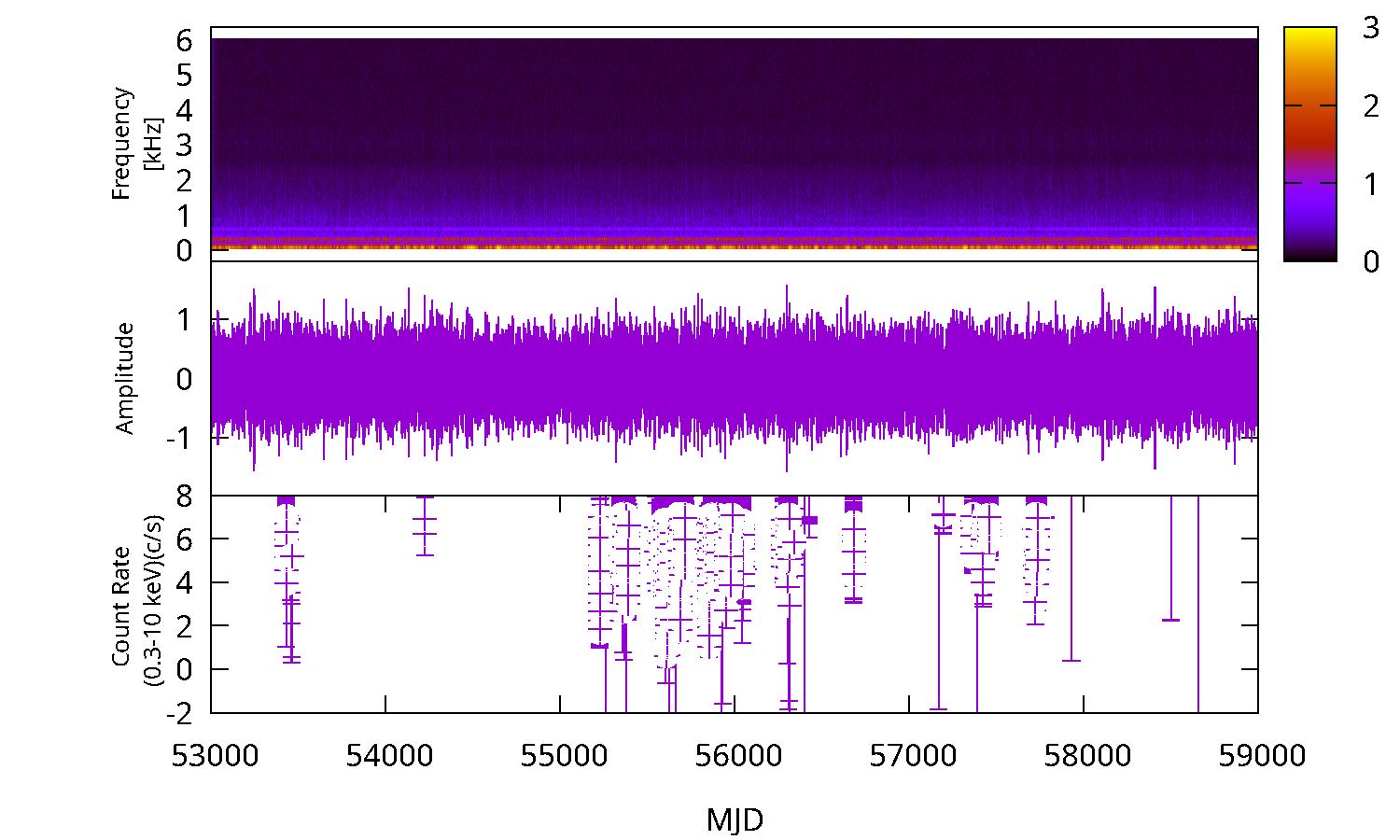}
  \includegraphics[width=8.0cm]{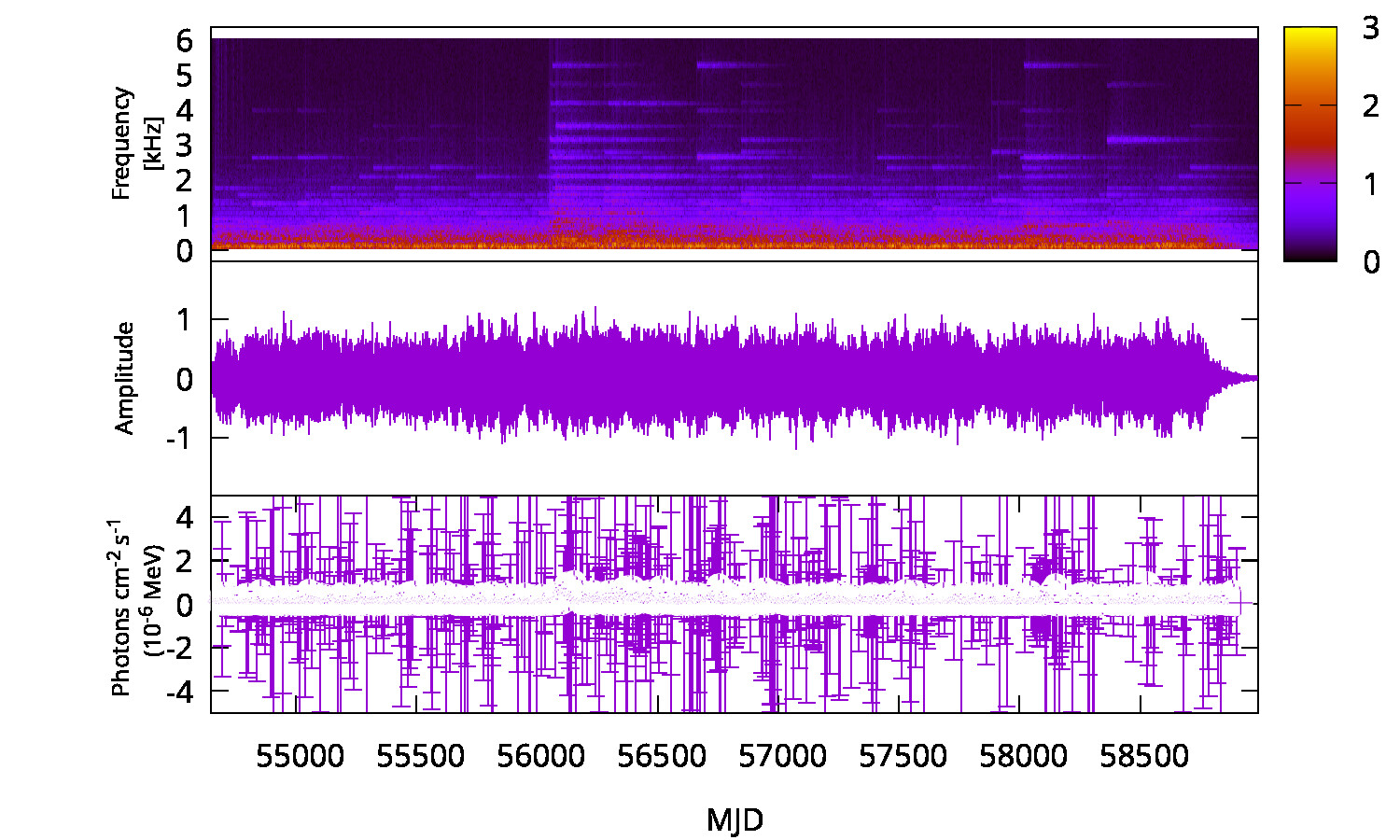} \\
  \end{center}
\caption{From top to bottom, left to right, the Figure shows panels
	of optical, X-ray and \( \gamma \)-rays  light curves,
	waveforms of the sonification as a function of time, and
	spectrograms of the blazar Mrk~421. The sonification is
available in \url{https://www.guijongustavo.org/datasonification/mrk421/playlist.html}}.
\label{fig:wave-mrk421}
\end{figure*}

%%%%%%%%%%%%%%%%%%%%%%%%%%%F I G U R E %%%%%%%%%%%%%%%%%%%%%%%%%%%%%%

%%%%%%%%%%%%%%%%%%%%%%%%%%%F I G U R E   W A V E F O R M  OJ287 %%%%%%%%%%%%%%%%%%%%%%%%%%%%%%
\begin{figure*}
  \begin{center}
  \includegraphics[width=8.0cm]{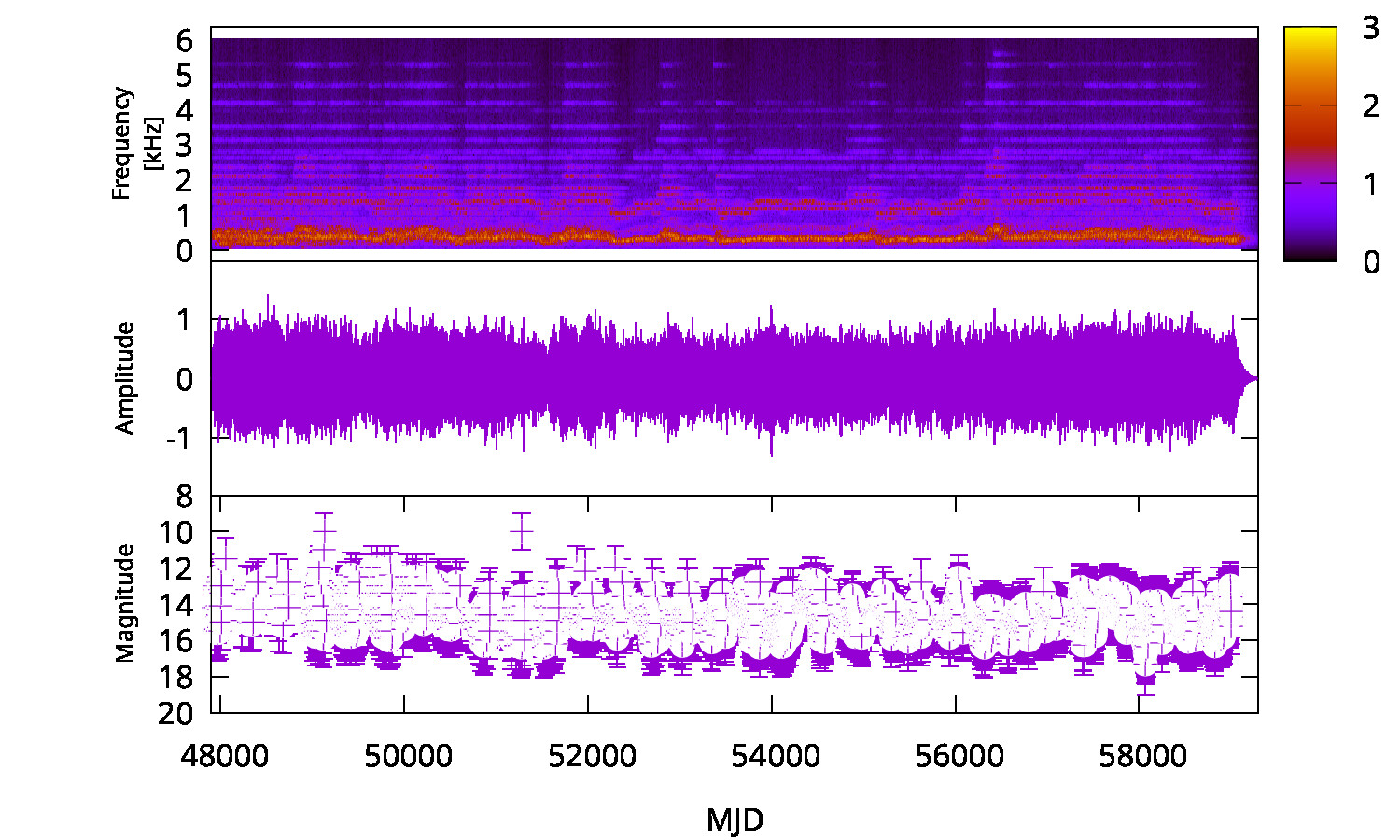} \\
  \includegraphics[width=8.0cm]{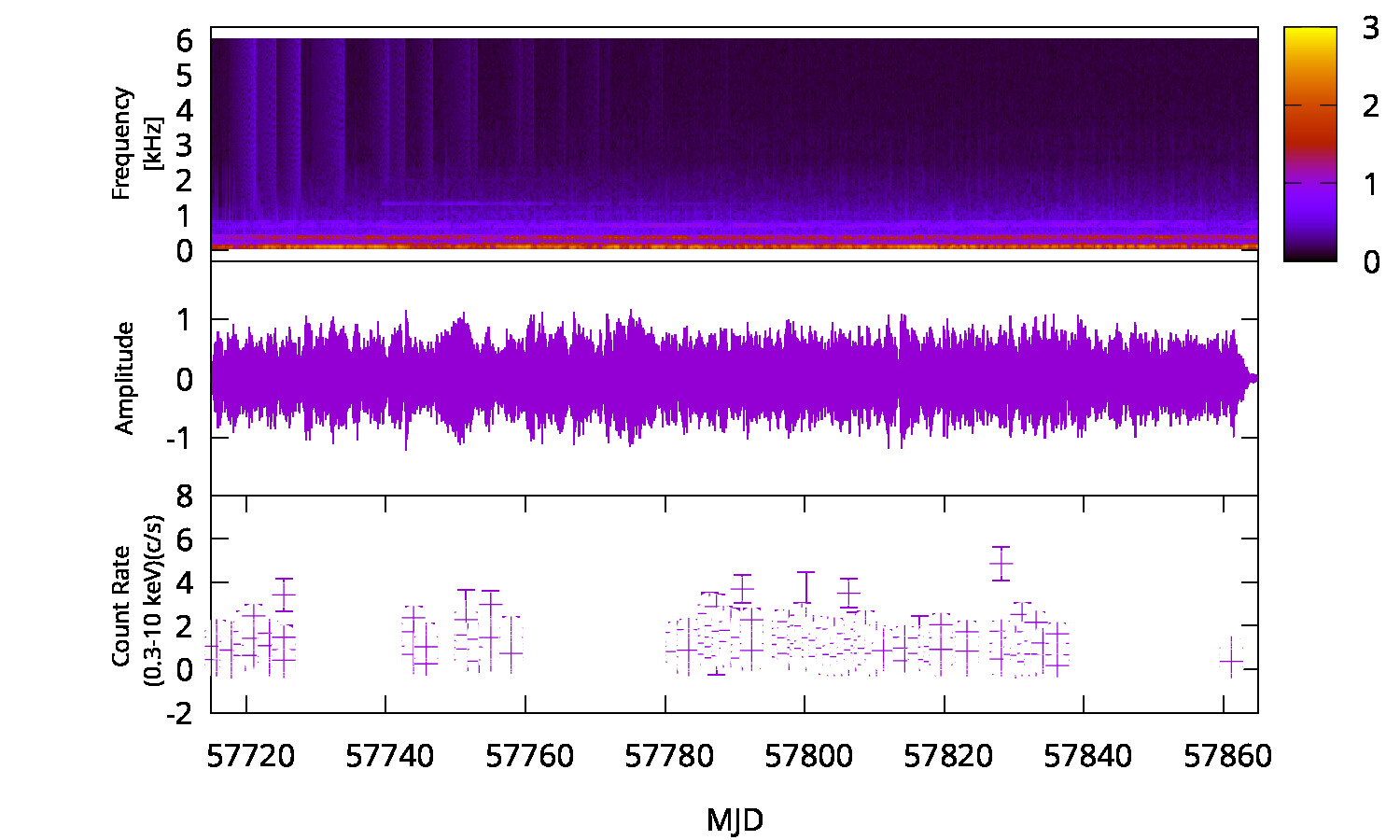}
  \includegraphics[width=8.0cm]{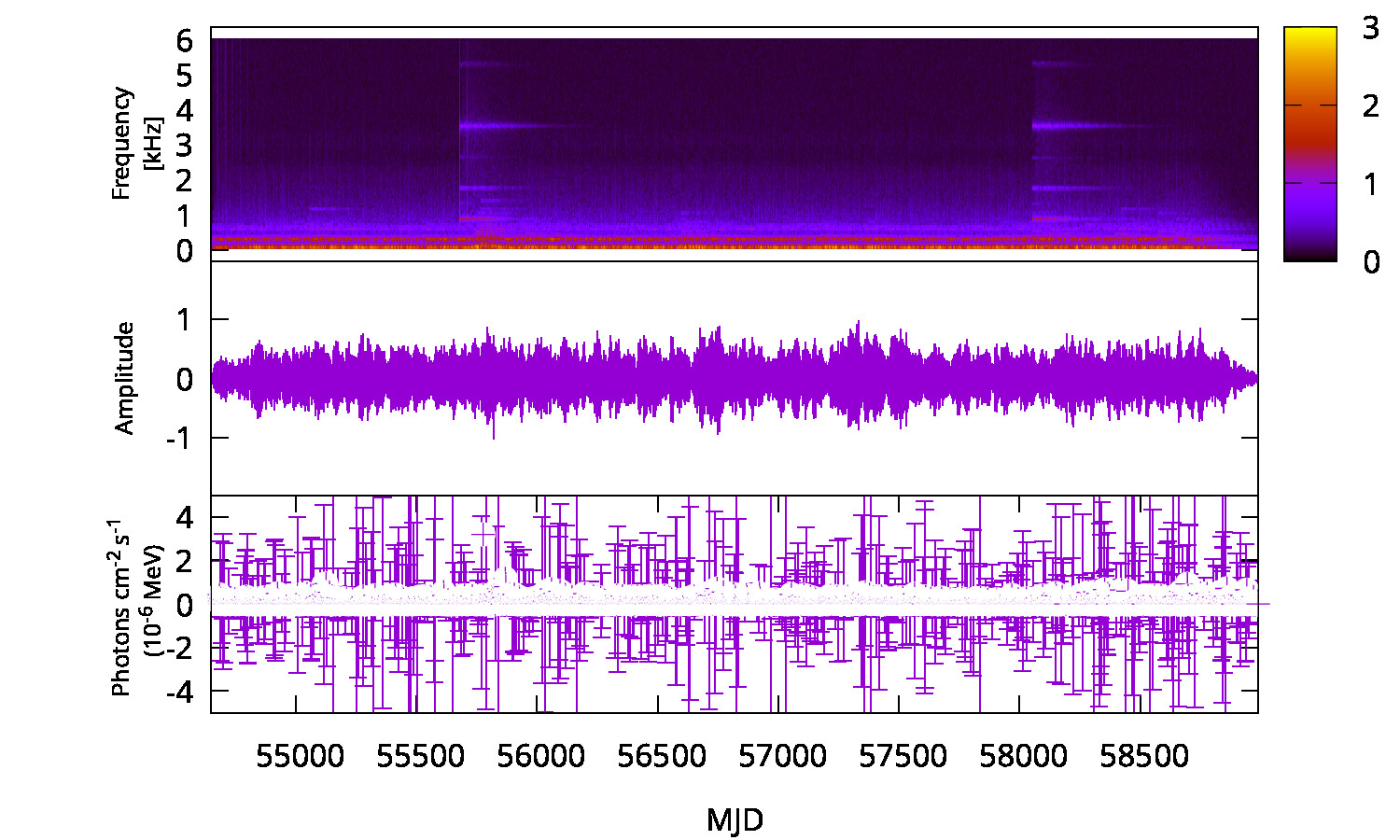} \\
  \end{center}
\caption{From top to bottom, left to right, the Figure shows panels
	of optical, X-ray and \( \gamma \)-rays  light curves,
	waveforms of the sonification as a function of time, and
	spectrograms of the blazar OJ~287. The sonification is
 available in \url{https://www.guijongustavo.org/datasonification/oj287/playlist.html}}.
\label{fig:wave-oj287}
\end{figure*}
%%%%%%%%%%%%%%%%%%%%%%%%%%%F I G U R E %%%%%%%%%%%%%%%%%%%%%%%%%%%%%%

\subsection{Production}
\label{produuction}
\noindent
Once the processed data were available, the production consisted of
converting signal to sound. 

For this purpose, each dataset was first normalized
using a linear min-max scaling:

\begin{equation}
\textrm{normalized value} = \frac{\text{flux} - \min(\text{flux})}{\max(\text{flux}) - \min(\text{flux})},
\end{equation}

where flux is the measured brightness at a given time,
and $\min(\text{flux})$ and $\max(\text{flux})$ correspond to the
minimum and maximum flux values in the dataset. This procedure ensures
that all data points are rescaled to the $[0,1]$ interval, facilitating
their subsequent mapping into musical notes. 

The normalized values were then mapped to Musical Instrument Digital
Interface (MIDI) notes using Parameter Mapping Sonification (PMSon) with
the `one-to-many' mapping technique~\citep{grond2011parameter}. In this
approach, each normalized data point triggers a primary note (encoding
flux as pitch) and a set of characteristic overtones, producing a richer
auditory representation. Additionally, the amplitude (loudness) of each
note is directly tied to the flux, meaning that the same physical quantity
is represented both in pitch and amplitude. This dual encoding enhances the
perception of variations, trends, and transient events in the light curves.

To map the normalized values to the notes of the scale, the interval
[0, 1] was divided into seven equal bins, each corresponding uniquely to
one of the seven notes of the D minor scale within a single octave. For
example, a normalized value of 0.0 corresponds to the first note (D),
a value of 0.5 corresponds to the fourth note (G), and a value of 1.0
to the highest note (C) in the octave.

The sonification was implemented in the C3 octave, with a tempo of 80
beats per minute (bpm), where each note represents one time step of the
light curve. In this context, the 80 bpm provides a temporal spacing
between successive notes, corresponding to the time evolution of the
observed flux. Different octave ranges were assigned to each wavelength
band during post-production. This choice does not imply any physical
relationship between electromagnetic wavelength and pitch; rather, it was
a perceptual design decision intended to enhance audibility and facilitate
differentiation between bands in the multiwavelength sonification.

Because the data span different time intervals and have varying sampling
rates across wavelength bands, the light curves were rescaled in time
for the sonification. Successive data points were mapped to fixed note
durations at 80 bpm, representing the relative temporal evolution of each
dataset. Absolute timing (MJD) is not preserved in the sonification;
instead, temporal structures and patterns within each wavelength are
maintained. This allows auditory perception of variability in each band,
while the differences in time span and sampling between bands provide
complementary insights. Key intervals, including known quasi-periodicities,
are annotated in the figures and on the project web page to guide the
listener in interpreting the sonified data.

The programming was carried out using the MIDItime 1.1.3
package~\citep{Mark2017} in \texttt{Python 3}. Data uncertainties were
not directly encoded in the sonification, although they informed the
preprocessing stage by guiding the removal or correction of anomalous
points prior to mapping.

\subsection{Post-production}

With the MIDI instrument interface of the Linux MultiMedia Studio (LMMS)
software, MIDI notes were synthesised using the ZynAddSubFX plugin. This
plugin is an open source synthesizer that combines different methods
of audio synthesis: additive synthesis, subtractive synthesis, Fourier,
and an algorithm for generating wavetable synthesis. The sound used to
represent the MIDI pulses was a tinkle bell musical instrument without
effects.

The tinkle bell produces a percussive, bell-like sound
with a fundamental pitch and a set of inharmonic overtones (i.e., partials that are not exact integer multiples of the fundamental frequency). Each
note has a rapid attack and an exponential decay determined by the
instrument's envelope. The spectral components scale proportionally with the input pitch, so higher notes produce perceptually brighter sounds. This
instrument was chosen because its clear and transient timbre and distinctive spectral structure allow
individual data points to be audibly distinguished, even in dense
sequences.

Parameter Mapping Sonification (PMSon) was applied using the `one-to-many'
mapping technique. In this approach, a single normalized data point
generates multiple simultaneous partials based on the tinkle bell's
harmonic structure. Specifically, each normalized data value is first
mapped to a primary MIDI note; this note then triggers the instrument’s
characteristic overtones, producing a richer auditory representation and
expanding the perceptual pitch content. To ensure perceptual separation
between wavelength regimes (radio, optical, X-ray, \(\gamma\)-ray),
each band was assigned to a distinct octave range via pitch offsets
applied after the mapping stage. This strategy preserves the relative
variations within each light curve while allowing the different regimes
to occupy clearly distinguishable pitch regions.

Each note has a constant duration of one beat at 80 bpm, corresponding to the temporal spacing defined in the sonification. The tinkle bell instrument was configured with an exponential decay of approximately 0.5 seconds, allowing individual notes to be clearly perceived while avoiding excessive overlap. Temporal overlap naturally occurs if data points are closer than the beat interval. Gaps in the light curves are represented by silent intervals of corresponding length, preserving the temporal structure of the original data.

The multifrequency sonified objects are illustrated in figures showing: 
\begin{enumerate}
	\item the original light curve (time series),
	\item the sonified waveform, and 
	\item the corresponding spectrogram, allowing visualization of both temporal and frequency content. 
\end{enumerate}

The auditory results of these sonifications can be accessed at the \href{https://www.guijongustavo.org/datasonification}{Datasonification Project} website. 

\subsubsection{Waveforms and spectrograms}

The visual representation of the objects were represented in three ways:
the original light curves, the sonification waveforms and the spectrograms. All plots are aligned along the time axis for direct comparison.

For each blazar, the waveforms represent the audio
signal as amplitude versus time, where amplitude corresponds to the
perceived loudness of each note. Since the datasets were normalized
prior to sonification, the maximum amplitude in all waveforms is 1.
Importantly, the waveforms do not encode pitch; they only show how the
loudness of the sound varies over time.

The spectrograms were generated from the sonified audio
rather than from the original light curve data. They display the
distribution of frequencies in the sonification as a function of time.
The vertical color bar indicates the intensity of the signal in each
time–frequency bin. Warmer colors (e.g., red or yellow) correspond to
stronger signal components, while cooler colors (e.g., blue or black)
indicate weaker ones. The presence of multiple bright frequencies within
a single time bin arises both from the harmonic overtones characteristic
of the tinkle bell instrument and from temporal overlap of successive
notes.

Together, the light curves, waveforms, and spectrograms
provide complementary perspectives on the same data: the original flux
variations, the time-domain structure of the sonification, and the
distribution of its pitch components across time.

\section{Discussion}
\label{discussion}

\noindent
Using a bell as a sonification instrument in this part
of the production enhances clarity and auditory distinction.
Its resonant tone, rich in overtones, has been shown to
facilitate the auditory salience of events by making them stand out from
the background \citep{kramer1999sonification, grond2011parameter}. Previous
studies in auditory display and sonification highlight that instruments
with clear attack and sustained resonance improve the perceptual
separation of events, reducing masking effects and allowing listeners to
identify patterns more easily \citep{hermann2008taxonomy}. Therefore, the tinkle bell timbre was
chosen to ensure that important variations in the data could be
perceived without blending with other sounds.

Additionally, its sharp attack (well-defined onset) allows for precise
marking of specific points in time, which is particularly useful for
sonifying time series or discrete events. The bell
timbre is also characterized by a rich harmonic spectrum: when a note is
struck, besides the fundamental frequency, several overtones of
different amplitudes are produced. These overtones increase the
perceptual brightness and richness of the sound, which improves the
listener's ability to discriminate subtle variations between consecutive
notes \citep{zwicker1999}. In practice, this means that small changes
in the mapped data values are more easily perceived as distinct auditory
events, as the harmonics reinforce pitch perception and provide
additional auditory cues beyond the fundamental tone.

For example, in the light curves presented here, 
the amplitude of each note is directly tied
to the flux value of the corresponding data point: higher flux
corresponds to a higher amplitude. The note length is constant for all
points and is chosen to ensure that each note sufficiently covers the
temporal spacing between consecutive measurements, effectively bridging
gaps in the light curve. This approach allows abrupt changes in
intensity or volume to be perceived by ear, even if they are not
visually evident in the waveform or the original light curve.
Consequently, sudden increases or decreases in flux are more perceptible
auditorily, providing an additional modality for identifying significant
variations in the data.

This is the advantage of sonification: features that may
be difficult to perceive visually, such as rapid short-term flares or
subtle oscillations in the flux, can be detected by the ear. When
sonified, these variations correspond to sudden changes in pitch or
amplitude, allowing listeners to perceive patterns and events that might
escape visual inspection of the light curves.

Synchronising the three plots (light curve, waveform, and spectrogram)
along the time axis provides a multidimensional representation of the
data. The light curve illustrates how the intensity or flux of the signal
changes over time, offering a global view of the source's luminosity or
intensity. This makes it possible to identify transient events such as
peaks, dips, eclipses, explosions, or periodic variations.

The waveform captures the full temporal evolution of the signal,
providing a zoom into temporal variations that might not
be visible in the light curve. For example, the oscillations in Mrk~501
that appear as small irregularities in the light curve allow us to
separate short-term variations from instrumental or observational noise.
Similarly, the sonification and waveform visualization make it possible
to correlate the peaks in the light curve with their amplitude and
timing, such as the 224-day periodicity observed in
Mrk~501~\citep{gustavo}, highlighting both the intensity and the temporal spacing.

The spectrogram is computed from the sonified audio signal rather than
from the original astronomical data. It decomposes this audio waveform
into its frequency components, showing how the pitch content generated
by the sonification evolves over time. For example, in the case of PKS
J2134-0153 (Figure~\ref{fig:wave-J2134-0153}), short-lived X-ray and
\(\gamma\)-ray flares that appear only as subtle bumps in the light curve
emerge as transient bright bands in the spectrogram. Because the tinkle
bell instrument produces a fundamental pitch together with inharmonic
overtones, each mapped data event spreads across multiple frequency
bands, which become visible in the spectrogram. This visualization
therefore reveals the temporal distribution, harmonic structure, and
overlap of sonified events, complementing both the light curve and the
auditory perception.

Combining these three plots provides a multiscale analysis:
\begin{itemize}
\item The light curve offers a general view of the signal's behaviour
over time. For example, in OJ~287, the quasi-periodic variability is visible is visible in the light curve as a series of recurrent peaks, establishing the large-scale context of variability.
\item The waveform adds fine detail about amplitude variations.
Rapid flares that appear only as small bumps in the light curve become clearly distinguishable in the waveform, allowing us to separate genuine short-term variability from instrumental noise.
\item The spectrogram reveals frequency properties, uncovering patterns
not evident in the other two. In the case of Mrk~421, the use of the tinkle bell instrument generates harmonics that appear as multiple bright bands in the spectrogram. While this complicates the auditory perception, it enhances the visual identification of event timing and harmonic structure, making subtle features—such as overlapping flares—more apparent.
\end{itemize}

This integrated approach enables the correlation of complex events. For
instance, a peak in the light curve can be examined in the waveform
to determine if it represents a sudden event and in the spectrogram to
check for associated frequency changes. This is relevant for identifying
specific physical phenomena, such as periodic pulses, spectral bursts, or
emission changes, often observed in the study of blazars or binary
stars. The corresponding spectrogram of Mrk~501, for example, shows
transient bright frequency optical band, indicating the presence of
oscillations.

Moreover, this combination is key to differentiate real signals
from noise. An apparent change in the light curve that turns out to be
noise can be confirmed using the waveform and spectrogram if no clear
patterns in amplitude or frequency are detected. For transient events or
explosions, the light curve can measure the event's total duration and
peak intensity, the waveform provides details about the pulse dynamics,
and the spectrogram highlights the specific frequency bands involved.

For instance, in case of OJ~049 a small fluctuation in the light curve
initially appears as a possible flare. However, inspection of the
waveform shows no coherent pulse dynamics, and the spectrogram lacks any
associated frequency bands, confirming that the fluctuation is
consistent with observational noise. The light curve reveals the overall
flare duration and peak intensity, the waveform captures the sharp rise
and decay dynamics, and the spectrogram highlights transient frequency
components produced by the sonified signal.

This integrated analysis of OJ~049 not only confirms
which apparent flares are artifacts of noise but also reveals genuine
variability signatures. In particular, during epochs of increased
activity, the waveform displays brief amplitude surges that correspond
to microflares lasting only a few hours, while the spectrogram shows
simultaneous enhancements across multiple frequency bands. These
correlated features indicate that the variability arises from real
physical changes-most likely turbulent processes or shock propagation
within the jet-rather than from instrumental or sampling effects. Thus,
combining the three representations enables a clearer discrimination
between stochastic noise and intrinsic variability in the blazar's
emission.

The synchronised combination of light curve, waveform, and spectrogram
provides a comprehensive and integrated view of the data: intensity,
temporal dynamics, and frequency structure. This enhanced analysis,
facilitates the identification of complex events and improves the
interpretation of physical phenomena. In
the case of OJ~049, the joint examination of the three representations
reveals that apparent fluctuations in the optical light curve without
corresponding features in the waveform or spectrogram are consistent
with observational noise. Conversely, when a true flare occurs, the
waveform shows coherent amplitude rises and decays, and the spectrogram
displays transient frequency bands linked to the event. Similarly, in
Mrk~421, correlated peaks across both optical and X-ray spectrograms
highlight multiwavelength variability episodes, confirming that the
combined visual–auditory approach enhances the detection and
understanding of real astrophysical processes.

As mentioned in the Introduction, there are already several
works on data sonification in astronomy, which explore different mapping
strategies and applications. For instance, \citet{Dubus2013} provided a
systematic review of sonification mapping strategies for physical
quantities, establishing a foundation for perceptual and analytical
approaches. More recent studies, such as \citet{Guiotto2024}, evaluated
the effectiveness of sonification for time-series data exploration,
highlighting its potential to enhance data comprehension beyond visual
analysis. Likewise, \citet{Tucker2022} tested the efficacy of
sonification for signal detection in light curves through the
\texttt{ASTRONIFY} framework, demonstrating measurable improvements in
identifying transient phenomena.

From a broader perspective, \citet{Misdariis2022}
examined how sound experts perceive astronomy sonification projects,
emphasizing both their scientific and communicative dimensions.
Similarly, \citet{Casado2024} discussed multimodal analysis approaches,
integrating auditory and visual representations to strengthen the
interpretative process of astrophysical data. These works not only
confirm the analytical value of sonification but also underscore its
pedagogical potential~\citep{zanella2022}, as it enables the public to experience complex
astronomical data in an engaging manner.

While previous works on sonification have focused
primarily on mapping static data or images, they generally do not
display how variations, trends, cyclicity, periodicity, or seasonality
evolve over time. In contrast, our approach focuses on the sonification
of temporal data -- specifically, light curves -- to capture the dynamic
behavior of astrophysical phenomena. 

We applied this method to nine active galaxies hosting
supermassive black holes with surrounding accretion disks and
relativistic jets, whose radiation exhibits multifrequency variability.
The sources analyzed include Mrk~501, Mrk~1501, Mrk~421, BL~Lacertae,
AO~0235+164, 3C~66A~(PKS~0219+428), OJ~049~(PKS~0829+046), OJ~287, and
PKS~J2134-0153. For each case, we present the sonified data alongside
their corresponding light curves, waveforms, and spectrograms, allowing
for a multiscale and multimodal interpretation of their emission
behavior.

\section{Conclusions}
\label{conclusions}
\noindent
This study presents a multifrequency exploration of blazars, combining
visualisation techniques (light curves, waveforms, and spectrograms)
with sonification as complementary tools for representing and analysing
their multifrequency variability. The integration of these methods not
only enables a richer interpretation of the data but also provides 
perspectives for scientific communication, particularly for visually
impaired communities.

In addition, these sonifications serve a dual purpose
for the visually impaired community. First, they provide a means
for BVI scientists to perceive patterns, periodicities, and transient
events directly through auditory cues, supporting independent data
exploration. Second, they can be used as an educational and
communicative tool, enabling the broader BVI audience to access and
understand complex astrophysical phenomena. For example, short-lived oscillations that appear as subtle fluctuations in a light curve can become more perceptually salient in the sonification, allowing listeners to detect features that are otherwise difficult to visualise.

Although this study does not aim to definitively
demonstrate periodicities, the auditory representation provides an
approach to detect potential patterns, correlate events
across different time scales, and explore the complex dynamics of these
astrophysical objects. By demonstrating how sonification can complement
traditional visualizations, this work establishes a methodological
foundation for future research, including systematic searches for
periodicities, transient events, and multiwavelength correlations in
astronomical data.

This multimodal approach enables a more inclusive and
diverse understanding of blazar behaviour by allowing data to be
explored both visually and auditorily. For example, visually impaired
(BVI) scientists can access the temporal dynamics, amplitude variations,
and frequency patterns in light curves through sonification, while
sighted researchers can combine waveforms and spectrograms with sound to
identify subtle features that might be missed visually. This integration
enhances accessibility and provides complementary perspectives on the
same data, improving the detection of flares, oscillations, and other
transient phenomena.

The radiation processes in the studied blazars are
multi-frequency, including synchrotron emission, inverse Compton
scattering, and thermal radiation from the accretion disk. These
processes produce variability in different energy bands (radio, optical,
X-ray, and gamma-ray), which we sonified. As a result, the listener can
perceive changes associated with specific radiative mechanisms: for
example, rapid optical flares generated by synchrotron emission
correspond to short, high-pitched notes, while slower variations in
gamma-ray flux appear as longer, lower-pitched tones. In this way, the
sonification provides an auditory representation of the physical
processes driving the emission, linking observable features to their
underlying astrophysical origins.

The graphs presented for each blazar are particularly
useful for understanding the sonification in visual terms. By pairing
the visualizations with the corresponding sounds, one can see how
changes in flux or luminosity are reflected in the amplitude (volume)
and frequency (pitch) of the sonified signal. This visual-auditory
correspondence aids in interpreting complex temporal patterns, verifying
features detected by ear, and providing a more comprehensive
understanding of the data.

The sonification of data allows the extraction of
substantial information, such as searching for periodicities, detecting
increases in power, identifying regularities, or noting absences of
data. Beyond visualizing the data, we sonified it to extend the
accessibility of our results to a broader audience. In particular, this
includes both the general public, who can gain an engaging and intuitive
understanding of astrophysical phenomena through sound, and blind or
visually impaired (BVI)~\citep{perez2019towards}. individuals, including both laypeople and
professional astronomers. By translating complex temporal patterns into
auditory cues, the sonifications provide a pathway for BVI audiences to
perceive and interpret features that might otherwise rely solely on
visual inspection.

\section*{Acknowledgements}
\label{acknowledgements}
% We thank an anonymous referee for the very useful comments 
% made to the manuscript.

This work was supported by a PAPIIT DGAPA-UNAM grants IN110522 and IN118325. 
GMG and SM
acknowledge support from  Secihti (378460, 26344).  We thank the variable star database
of observations from the AAVSO International Database contributed by
observers worldwide.  We thank the public data observations from the Swift
data archive and the Fermi Gamma-Rays Space Telescope collaboration for
the public database used in this work.

The operation of UMRAO is supported by funds from the University of
Michigan Department of Astronomy, which we acknowledge too. `This research
has made use of data from the University of Michigan Radio Astronomy
Observatory which has been supported by the University of Michigan
and by a series of grants from the National Science Foundation, most
recently AST-0607523'.

We thank  the use of archival calibrated VLBI data from the Astro-geo
Center database maintained by Leonid Petrov.

This work has used observations obtained with the Samuel Oschin 48-inch
Telescope at the Palomar Observatory as part of the Zwicky Transient
Facility project. ZTF is supported by the National Science Foundation
under Grant No. AST-1440341 and a collaboration including Caltech,
IPAC, the Weizmann Institute for Science, the Oskar Klein Center at
Stockholm University, the University of Maryland, the University of
Washington, Deutsches Elektronen-Synchrotron and Humboldt University,
Los Alamos National Laboratories, the TANGO Consortium of Taiwan,
the University of Wisconsin at Mil- waukee, and Lawrence Berkeley
National Laboratories. Operations are conducted by COO, IPAC,
and UW.  This work has made use of publicly available data from
ZTF (\url{https://irsa.ipac.caltech.edu/Missions/ztf.html}).

\section*{Data Availability}
The data underlying this article are available in the article and in
its online supplementary material, together with the complementary webpage:
\url{https://www.guijongustavo.org/datasonification}.

\appendix
\section{Blazar Sample and Specifications}
\label{apendice}
\noindent
For each dataset, the time resolution and integration per data point are
determined by the corresponding instrument or database. In the optical
band, ZTF data consist of single exposures of 30 s~\citep{Masci2019},
while AAVSO records typically correspond to integrations of 60–300 s
depending on the observer~\citep{Kafka2021}. In the radio band, UMRAO
data points were obtained from integrations of approximately 30 minutes
per observation~\citep{Aller1985ApJS}, and Astrogeo VLBI measurements
correspond to individual interferometric sessions~\citep{Petrov2021}.
For the X-ray band, Swift-XRT provides observations with effective
exposures ranging from a few hundred to a few thousand
seconds~\citep{Burrows2005}. In the case of the \(\gamma\)-ray band,
Fermi-LAT light curves are generally binned into one-week intervals to
ensure sufficient photon statistics~\citep{Ballet2023}. These values
represent the effective time resolution of the light curves used in this
work.

Each blazar selected had the following specifications:   

\textbf{(1) Mrk~501},  has a redshift \(z\) = 0.034 that corresponds to a
distance \(\sim\) 140 Mpc and R. A. = 16h 53m 52.2s, Dec. = +39° 45' 37''. 
It has been monitored in radio~\citep{Richards2011ApJ},
optical~\citep{smith2009}, X-rays~\citep{Abdo2011ApJb}, and
$\gamma$-rays~\citep{Dorner2017}. This object has several periodicities
reported~\citep{Bhatta2019}. Recently~\citet{magallanes2024} have reported a periodicity \(\sim\) 224 days. 

The light curve in optical was built
using the database from 1998 June 24 to 2021 September 12. The Swift
database is from 2008 October 2 to 2021 October 8. And the \( \gamma
\)-rays, covers a time interval from 2008 August 04 up to 2020 March 13.
Table~\ref{tab:all-blazars} shows specific information about the records
and the light curve is presented in Figure~\ref{fig:mrk501}.

\textbf{(2) Mrk~1501} is a FSRQ with redshift of \(z\) = 0.089338 
at a distance of \(\sim\) 377 Mpc~\citep{Sargent1970}. This blazar 
presents short-lived $\gamma$-ray emission flares~\citet{Arsioli2018A}.  
It is in the Pisces's constellation, with R. A. = 00h 10m, and Dec. = +10° 58'.~\citep{Markarian1989}.

For the optical, the dataset was taken from the ZTF database,
from 2010 July 25 to 2021 November 25. In the case of X-rays the database
comes from the Swift observatory from 2008 October 2 to 2021 March 22. For
the period of 2008 August 4 to 2022 October 22, the dataset is from the Fermi database. Table~\ref{tab:all-blazars} shows the summary of the dataset, and 
Figure~\ref{fig:wave-mrk1501} presents the light curves. 

\textbf{(3) BL Lacerta} was discovered in 1929 by~\citet{Hoffmeister1929}
and~\citet{Oke1974} measured its redshift \(z\) = 0.07 which corresponds
to a distance of \(\sim\) 0.276 Gpc. It is in the constellation Lacertae,
with R. A. = 22h 02m 43.3s, Dec. = +42° 16' 40''~\citep{Miller1978ApJ}.

The dataset in radio for this object was obtained from UMRAO from 
2009 September 1 to 2012 May 21. For optical, the dataset used was
from AAVSO in an interim of 1969 December 14 to 2022 March 13. For
X-rays, the time interval is 2008 October 2 to 2022 November 10, taken
from the Swift Observatory. And for \( \gamma \)-rays, the time of 
observation was from
2008 August 4 to 2022 March 11. Table~\ref{tab:all-blazars}
presents the summaries of the dataset and the light curves are shown in
Figure~\ref{fig:wave-bllac}.

\textbf{(4) AO~0235+164} is a BL~Lac object\footnote{\citet{D'Elia2015MNRAS}
suggest that it could classify as an FSRQ according to its strong H\(
\alpha \) feature in the IR band.} with a R. A. = 02h 35m 52.6s, Dec. =
16° 24' 05''.~\citep{spinrad1975} and redshift of \(z\) = 0.940 \(\sim\)
corresponding to a distance of \(\sim\) 3 Gpc~\citet{Cohen1987ApJ}.

For AO~0235+164, the optical dataset was obtained from
1998 August 25 to 2020 September 13, and the database used was from
AAVSO. For X-rays, the dataset used was from Swift from 2008 October 2
to 2017 December 7. The Fermi \( \gamma \)-rays database was taken
from 2008 August 4 to 2020 April 5. Table~\ref{tab:all-blazars} shows a
summary of these details and Figure~\ref{fig:wave-AO0235+164} presents 
the corresponding light curves.

\textbf{(5) 3C~66A} has a redshift of \(z\) = 0.34 corresponding
to \(\sim\) 1.4 Gpc~\citep{Torres-Zafra2018MNRAS}. This
BL~Lac object has been observed on optical
wavelengths~\citep[e. g.][]{Bottcher2005ApJ,Lainela1999ApJ}. It has R.A. =
2h 23m 12s and Dec.= 43° 0.'7 reported by~\citet{Errando2009arXiv}
in \( \gamma \)-rays.

The radio records from the Astro-geo VLBI were taken from 
1997 January 11 to 2020 June 4. For optical,
the dataset used was from AAVO from 1994 February 1 to 2022 February
10. The Swift X-rays database used is from 2008 October 2
to 2022 December 9. All these specifications are presented in 
Table~\ref{tab:all-blazars} and the light curves are shown in 
Figure~\ref{fig:wave-3C66A}.

\textbf{(6) OJ~049} has a redshift \(z\) = 0.18 \(\sim\), corresponding to
a distance of \(\sim\) 760 Mpc~\citep{Falomo1991AJ}. It was detected as
an X-rays source~\citep{della1990catalog} and as a GeV \( \gamma\)-rays
source~\citep{Fichtel1994ApJS, vonMontigny1995ApJ, Mattox1998ASPC,
Mukherjee1997ApJ}. Its R. A. = 08h 31m 48.8s and Dec. = \( 04^{\circ}
\) 29' 39''~\citep{Fiorucci1996}.

For this blazar the databases were AAVSO, Swift and Fermi. 
The corresponding time periods were from 2008 January 9 to 2020
June 23, 2002 October 9 to 2020 March 31, 2006 October 23 to 2021
December 31 and 2008 August 5 to 2020 April 6, respectively. 
Table~\ref{tab:all-blazars} shows all these details and their
corresponding  light curves are presented in Figure~\ref{fig:wave-OJ049}.

\textbf{(7) PKS J2134-0153 } has a redshift \(z\) = 1.285, corresponding to
a distance \(\sim\) 5 Gpc~\citep{Truebenbach_2017} with R. A. = 21h 34m 10.3095s and Dec. = \( -01^{\circ} \) 53' 17.238''~\citep{Fomalont2000ApJS}. 

For optical, the database is
from ZTF with interim from 2009 October 4 to 2010 August 14. In X-rays,
the dataset is from Swift from 2008 October 2 to 2014 October 8. For \(
\gamma \)-rays, the database is Fermi, with records from 2008 August 4
to 2022 October 22. All these details are presented in 
Table~\ref{tab:all-blazars}.   Figure~\ref{fig:wave-J2134-0153} shows the 
corresponding light curves.

\textbf{(8) Mrk~421} has a R. A. = 11h 04m 27 31s and Dec. = \(
38^{\circ}\) 12' 31.80'', located at a redshift \(z\) = 0.031, which
corresponds to a distance \(\sim\) 130 Mpc)~\citep{Ulrich1975ApJ}.
This object shows
rapid ﬂux and polarisation variability~\citep[e. g.,][]{Fraija_2017}.

For optical, the dataset is from AAVSO database, from 1984
June 22 to 2020 April 13. In the case of X-rays, the database used
was Swift, from 2005 March 1 to 2020 April 1. For \( \gamma \)-rays,
the database Fermi gave the dataset from 2008 June 5 to 2020 March
8. The summary of all this is presented in Table~\ref{tab:all-blazars}. The
correspond light curves are shown in Figure~\ref{fig:wave-mrk421}.

\textbf{(9) OJ~287} is a BL~Lacerta type quasar situated at a redshift of
\(z\) = 0.306 \(\sim\) 4 Gly \(\sim\) 1 Gpc with R. A. = 08h 54m 48.87s
and Dec. = +20° 06' 30.6''. This blazar is a candidate
to host a binary supermassive black hole~\citep{Valtonen2006ApJ} with
a periodicity of 12 years in its optical wavelength~\citep{Shi2007Ap}.

For optical, the database was obtained from
AAVSO, from 1987 September 26 to 2020 June 3. Swift provided the dataset of
X-rays from 2008 November 12 to 2020 April 28. In the case of \( \gamma
\)-rays the dataset is from Fermi, from 2008 April 4 to 2020 June 6. 
Table~\ref{tab:all-blazars} show the records of this blazar. 
Figure~\ref{fig:wave-oj287} presents the corresponding light curves.

\section{Tables}
\label{tablas}
\noindent
In Tables~\ref{tab:blazar-summary-short} and~\ref{tab:all-blazars}, we use the following conventions: a
Records refers to a single photometric or flux measurement in a given
band, typically corresponding to one integration by the instrument or
observer. The term Interim~[d] indicates the total time span covered by the
dataset, expressed as years, months, and days~[y, m, d] between the first and last
observation date, regardless of gaps. The column Total~[d] refers to
the number of distinct calendar days within that interval for which at
least one measurement was obtained.

\begin{table*}
\centering
\begin{tabular}{| l | c | c | c | c |}
\hline
Blazar & R.A. & Dec. & Redshift & Distance \\
\hline
Mrk 501 & 16h53m52.2s & +39°45'37'' & 0.034 & 140 Mpc \\
Mrk 1501 & 00h10m & +10°58' & 0.089338 & 377 Mpc \\
BL Lacerta & 22h02m43.3s & +42°16'40'' & 0.07 & 0.276 Gpc \\
AO 0235+164 & 02h35m52.6s & +16°24'05'' & 0.940 & 3 Gpc \\
3C 66A & 02h23m12s & +43°00'07'' & 0.34 & 1.4 Gpc \\
OJ 049 & 08h31m48.8s & +04°29'39'' & 0.18 & 760 Mpc \\
PKS J2134-0153 & 21h34m10.3095s & -01°53'17.238'' & 1.285 & 5 Gpc \\
Mrk 421 & 11h04m27.31s & +38°12'31.8'' & 0.031 & 130 Mpc \\
OJ 287 & 08h54m48.87s & +20°06'30.6'' & 0.306 & 1 Gpc \\
\hline
\end{tabular}
\caption{Summary of the nine blazars analyzed in this work, including equatorial coordinates (R.A., Dec.), redshift, and approximate distance.}
\label{tab:blazar-summary-short}
\end{table*}

\begin{table*}
\centering
\begin{tabular}{| l | r | c | c | c | c |}
\hline
Blazar & Band & Records & Interim & Total & Dates \\
\ & \ & \ [y, m, d] & [d] & [d] & \\
\hline
Mrk 501 & Optical & 11,849 & 23, 2, 17 & 8,441 & 1998-06-24 -- 2021-09-12 \\
        & X-rays  & 28,000 & 12, 7, 9  & 4,607 & 2008-10-02 -- 2021-10-08 \\
        & $\gamma$-rays & 4,199 & 11, 7, 9 & 4,239 & 2008-08-04 -- 2020-03-13 \\
\hline
Mrk 1501 & Optical & 1,255 & 11, 4, 0 & 4,141 & 2010-07-25 -- 2021-11-25 \\
          & X-rays  & 11,150 & 12, 5, 20 & 4,554 & 2008-10-02 -- 2021-03-22 \\
          & $\gamma$-rays & 5,146 & 14, 2, 18 & 5,192 & 2008-08-04 -- 2022-10-22 \\
\hline
BL Lacertae & Radio & 413 & 12, 5, 25 & 4,560 & 2009-09-01 -- 2012-05-21 \\
            & Optical & 10,117 & 52, 2, 27 & 19,082 & 1969-12-14 -- 2022-03-13 \\
            & X-rays & 100,418 & 14, 1, 8 & 5,152 & 2008-10-02 -- 2022-11-10 \\
            & $\gamma$-rays & 4,892 & 13, 7, 7 & 4,967 & 2008-08-04 -- 2022-03-11 \\
\hline
AO 0235+164 & Optical & 196 & 22, 9, 17 & 8,055 & 1998-08-25 -- 2020-09-13 \\
            & X-rays & 61,146 & 9, 2, 5 & 3,353 & 2008-10-02 -- 2017-12-07 \\
            & $\gamma$-rays & 4,223 & 11, 8, 1 & 4,262 & 2008-08-04 -- 2020-04-05 \\
\hline
3C 66A & Radio & 101 & 23, 4, 22 & 8,545 & 1997-01-11 -- 2020-06-04 \\
       & Optical & 8,470 & 28, 0, 9 & 10,236 & 1994-02-01 -- 2022-02-10 \\
       & X-rays & 3,163 & 14, 2, 7 & 5,181 & 2008-10-02 -- 2022-12-09 \\
       & $\gamma$-rays & 4,223 & 11, 8, 1 & 4,262 & -- \\
\hline
OJ 049 & Optical & 76 & 17, 5, 22 & 6,383 & 2008-01-09 -- 2020-06-23 \\
       & X-rays & 16,111 & 15, 2, 8 & 5,548 & 2002-10-09 -- 2020-03-31 \\
       & $\gamma$-rays & 4,196 & 11, 8, 1 & 4,262 & 2008-08-05 -- 2020-04-06 \\
\hline
PKS J2134-0153 & Optical & 73 & 0, 10, 10 & 314 & 2009-10-04 -- 2010-08-14 \\
               & X-rays & 4,123 & 6, 0, 6 & 2,197 & 2008-10-02 -- 2014-10-08 \\
               & $\gamma$-rays & 4,223 & 11, 8, 1 & 4,262 & 2008-08-04 -- 2022-10-22 \\
\hline
Mrk 421 & Optical & 18,167 & 35, 9, 20 & 13,079 & 1984-06-22 -- 2020-04-13 \\
        & X-rays & 160,346 & 15, 1, 0 & 5,510 & 2005-03-01 -- 2020-04-01 \\
        & $\gamma$-rays & 4,129 & 11, 9, 3 & 4,294 & 2008-06-05 -- 2020-03-08 \\
\hline
OJ 287 & Optical & 13,754 & 32, 8, 6 & 11,939 & 1987-09-26 -- 2020-06-03 \\
       & X-rays & 9,464 & 11, 5, 16 & 4,185 & 2008-11-12 -- 2020-04-28 \\
       & $\gamma$-rays & 4,410 & 12, 2, 2 & 4,446 & 2008-04-04 -- 2020-06-06 \\
\hline
\end{tabular}
\caption{Summary of all electromagnetic bands studied for the nine
blazars. Records corresponds to a single flux measurement.
Interim~[d] is the total time span covered by the observations,
expressed in years, months, and days~[y, m, d]. Total~[d] indicates the
number of distinct days within the interim for which at least one
measurement is available.}
\label{tab:all-blazars}
\end{table*}


\bibliographystyle{mnras}
\bibliography{magallanes-mendoza}

%%%%%%%%%%%%%%%%%%%%%%%%%%%%%%%%%%%%%%%%%%%%%%%%%%

% Don't change these lines
\bsp	% typesetting comment
\label{lastpage}
\end{document}